\documentclass[a4paper,12pt]{article}
\usepackage[utf8]{inputenc}
\usepackage[affil-it]{authblk}
\usepackage{amsmath}
\usepackage{stackrel}
\usepackage{hyperref}
\usepackage{amssymb}
\usepackage{verbatim} 
\numberwithin{equation}{section}
\usepackage{url}
\usepackage{enumerate}
\newtheorem{theorem}{Theorem}[section]

\newtheorem{definition}{Definition}[section]
\title{A hyperbolic theory of relativistic conformal dissipative fluids}
\author{Luis Lehner${}^{1}$, Oscar A. Reula${}^{2,3}$ and Marcelo E. Rubio${}^{1,2,3}$}

\affil{
${}^{1}$Perimeter Institute for Theoretical Physics, Waterloo,\\ Ontario N2J 2W9, Canada\\
${}^{2}$Facultad de Matemática, Astronomía, Física y Computación,
Universidad Nacional de Córdoba, Ciudad Universitaria,\\
(5000) Córdoba, Argentina\\
${}^{3}$Instituto de Física Enrique Gaviola, CONICET, Córdoba, Argentina
}

\begin{document}

\maketitle

\begin{abstract}
We develop a complete description of the class of conformal relativistic dissipative fluids of divergence form, following the formalism described in \cite{geroch1990dissipative} and \cite{pennisi87}. This type of theories is fully described in terms of evolution variables whose dynamics is governed by total divergence-type conservation laws.

Specifically, we give a characterization of the whole family of conformal fluids in terms of a single master scalar function defined up to second order corrections in dissipative effects, which we explicitly find in general form. This allows us to identify the equilibrium states of the theory, as well as to derive constitutive relations and a Fourier-like law for the corresponding first-order theory heat flux. Finally, we show that among this class of theories-- and near equilibrium configurations-- there exist symmetric hyperbolic ones, implying that for them one can
define well posed initial value problems.
\end{abstract}

\newpage

\tableofcontents

\newpage

\section{Introduction}
\label{sec1}

Throughout a broad spectra of physics, relativistic fluids play a crucial role to describe the behavior of matter
and energy. In astrophysics, they arise as a description of compact stars and accretion disks \cite{VanRiper79,Zel'dovich71,Peebles80,anile1989relativistic}; in particle physics they are called for describing quark-gluon plasmas produced in high energy collisions (\cite{Barz85,CLARE1986177} and \cite{Jaiswal16} for a nice review); and  in cosmology they play a crucial role in describing the radiation--dominated epoch in the early universe (see the book \cite{mukhanov2005physical} and references therein). The understanding of its mathematical structure dates back to the early works of Lichnerowicz \cite{Lichne65,Lichne66}, who first proposed a relativistic theory of fluids in the late 60's, providing some results of existence and uniqueness of 
relevant solutions.

Physically, fluid dynamics can be derived as an effective description of interacting quantum fields when considering fluctuations that are of sufficiently long wavelength.  This scheme requires a kinetic theory approach, providing the corresponding precise definitions and relations among thermodynamic variables \cite{Betz:2009zz,Van:2011yn,calzetta2010linking,PhysRevLett.105.162501,0305-4470-28-23-033}. The resulting equations govern the
behavior of macroscopic properties of fluids in terms of a piecewise continuous density, velocity, and pressure, as well as relationships among them.

From a mathematical point of view, fluids are generally described
by different families of evolution equations, depending on the type of phenomena one is interested in \footnote{Such evolution equations could be or not of ``conservation laws". While this type of equations are useful for describing shocks during evolution, there may exist more general causal theories of dissipative relativistic fluids (see \cite{geroch1991causal}).}. In
the non-relativistic regime, for instance, perfect/(incompressible viscous) fluids are described by the so called \textit{Euler}/(\textit{Navier--Stokes}) equations respectively. While a relativistic version of Euler's equation is readily
obtained, the inclusion of dissipative effects in relativistic regimes\footnote{And, more generally, the issue of finding globally regular solutions of the corresponding evolution equations.} is non trivial, and introduces several subtleties already at the formal level. The basic reason for this is that a Galilean description of fluids --that yields the Navier--Stokes equations, results in \textit{parabolic} equations, as the \textit{diffusion equation}. Its nature accounts for perturbations propagating at infinite speeds and thus cannot be accommodated into the causal structure imposed by relativity \cite{Hiscock:1985zz,nagy1997exponential,kreiss1997global}. This is evident, for instance, in the early attempts of relativistic fluids due to Eckart and Landau-Lifschitz \cite{landau2013fluid}, in which they considered as basic dynamical variables the four-velocity of the fluid, as well as some thermodynamical quantities. It became clear that a substantial enlargement of the system was needed in order to resolve the tension between phenomenological behavior --which favors a description à la Navier-Stokes-- and causality --which requires finite propagation speeds and so \textit{hyperbolic} systems \cite{geroch2001hyperbolic}.

As mentioned, a relativistic generalization of the hydrodynamic equations is not straightforward. There have been several proposals for suitable attempts, and all of them require accounting for new parameters beyond those encountered in the Navier--Stokes setting. In many cases, those parameters are difficult to estimate from available experiments, but also are thought to be mostly irrelevant in describing much of the phenomenology of interest. In particular, some attempts extend the set of dynamical variables to be the total stress-energy tensor of the fluid and the associated conserved particle number current. One
of them, proposed by Liu, Müller and Ruggeri \cite{liu1986relativistic}, contemplates total divergence-type equations, which render the issues of hyperbolicity and causality more transparent, and was extended later on \cite{muller2013extended}. A few years later, Geroch and Lindblom \cite{geroch1990dissipative} adopted this approach, but relaxing previously imposed symmetries (adopted without clear physical motivations, see \cite{geroch1991causal} and \cite{geroch1995relativistic}).

In this work we extend this approach in order to characterize and analyze the initial value problem of the general class of {\em relativistic conformal fluids}. The description shall be centered in the determination of a generating function, i.e., a scalar field defined on the whole spacetime, that contains all the information of the fluid, considering dissipative effects up to second order. This generating function will be determined by imposing conformal invariance to the corresponding dynamical equations, and in particular, the second order contribution will be crucial in order to guarantee well posedness of the theory near equilibrium states.

\subsection{Conformal fluids}
\label{conf-fluids}

As previously stated, one of the fundamental properties of relativistic fluid theories is that they represent the low energy limit of almost any quantum field theory. In particular, \textit{conformal} field theories lead to \textit{conformal} fluid theories at low energies \cite{bhattacharyya2008nonlinear}. This sort of fluids has the particular characteristic that they are \textit{conformally invariant}; i.e., dynamical equations are invariant under conformal transformations of the spacetime metric (sometimes known as Weyl transformations). 
Though certainly the assumption of 
conformal symmetry restricts the type of fluids under
consideration, it has been pointed out that in non--conformal fluids such
symmetry might emerge in suitable regimes \cite{boffetta12}.

Moreover, beyond the intrinsic importance of understanding the behavior of conformal fluids (and
their message to non--conformal ones), 
they have recently been linked to an a priori separate branch of
physics {\em geometry}. Indeed, the fluid/gravity correspondence \cite{Bhattacharyya:2008mz,Rangamani:2009xk,Ambrosetti:2008mt,Bredberg2012,1126-6708-2008-05-087}
elucidated a direct relation between a class of (perturbed) black holes in
asymptotically Anti de Sitter spacetimes in $d + 1$ dimensions and conformal
relativistic hydrodynamics in $d$ dimensions. For instance, for any solution
to the conformal fluid equations, it is possible to construct a black-brane
solution in one more dimension, allowing thus to explore a wide range
of aspects of black hole structure and their stability in terms of the corresponding fluid dual theory –and vicerversa.

\subsection{Overview, conventions and a notation list}
\label{overview}

This work is organized as follows. In section \ref{sec2} we state some generalities regarding divergence--type fluid theories, as well as particularize to the case of conformal theories. This shall be the theoretical set up in which the work is framed. Section \ref{sec3} is devoted to review conformal invariance, as well as introduce the notion of conformal weights. The next three sections contain one of the main results of the paper: a full characterization of
conformal fluids in terms of a single scalar function including dissipative effects. In section \ref{perfect-fluids} we discuss conformal perfect fluids and derive the equilibrium states of the theory, as well as sections \ref{1storder-section} and \ref{sec6} are dedicated to discuss in detail the contributions to the theory at first and second order in dissipation, respectively. In section \ref{sec7} we study 
the initial value problem of the full theory (that is, up to second order in dissipation), using the algebraic tools for hyperbolicity provided by Friedrichs, Lax and Geroch. We state and show the main theorem about near equilibirum symmetric hyperbolicity of the full theory, which constitute the second main result of this work. Final comments and a general conclusion is displayed in section \ref{sec8}.

Throughout this work, we consider a time-oriented background spacetime $\mathcal{M}$ with arbitrary dimension, $d$. We will adopt the signature $(-,+,+,+,\cdots)$ and denote spacetime indices with Latin lowercase letters $a,b,\cdots$. Latin uppercase indices $A,B,\cdots$ and Greek lowercase indices $\alpha, \beta, \cdots$ will be respectively reserved for the equations and dynamic fields vector spaces within Geroch's covariant formalism \cite{geroch1996partial}. Finally, natural units $G = c = k_B = 1$ will be assumed, where $G$ is Newton's universal constant, $c$ is the speed of light in vacuum and $k_B$ the Boltzmann's constant. 

\subsubsection{Definitions and notation}
\label{list-notation}

We present here a short list with some of the notation and definitions we shall use throughout this work (see Table 1). Of course, the same definitions shall be explicitly stated on each one of the sections in which they are used. Here we collect them in a unique list of definitions in terms of a pair or variables $(\xi_a, \xi_{ab})$ that will be adequately introduced in section 2. Finally, indicies of all tensor fields on $\mathcal{M}$ are raised and lowered using the background metric.

\begin{table}[h]
\begin{center}
\begin{tabular}{|c|c|}
\hline \textit{Quantity} & \textit{Definition} \\ 
\hline \hline
$g_{ab}$     & background metric              \\ \hline
$d$     & spacetime dimension               \\ \hline
$D_j$     & $\frac{d}{2} + j$, $j=0,1,2,\cdots$               \\ \hline
$\mu$     & $\xi^c \xi_c$               \\ \hline
$D$     & $\nabla_c \xi^c$               \\ \hline
$\nu$     & $\xi^{ab} \xi_a \xi_b$               \\ \hline
$\psi_1$  & $\xi^{ab}\xi_{ab}$               \\ \hline
$\psi_2$     & $\xi^{ab}\xi_b\xi_{ac}\xi^c$               \\ \hline
$\psi_3$     & $\nu^2$               \\ \hline 
$\dot{\mu}$     & $\xi^a\nabla_a \mu$               \\ \hline
$\dot{\nu}$     & $\xi^a\nabla_a \nu$                \\ \hline
$u^c$     & $\frac{\xi^c}{\sqrt{-\mu}}$               \\ \hline
$\ell^a$     & $\xi^{ab}\xi_b$               \\ \hline
$\dot{\ell}^a$     & $\xi^c \nabla_c \ell^a$               \\ \hline
$r^c$     & $\ell^c - \frac{\nu}{\mu}\xi^c$               \\ \hline
${\dot{\xi}}^c$     & $\xi^a\nabla_a\xi^c$               \\ \hline
$a^c$     & $u^a\nabla_a u^c$            \\ \hline
${\dot{\xi}}^{ab}$     & $\xi^c\nabla_c \xi^{ab}$     \\    \hline                  
\end{tabular}
\end{center}
\caption{List of definitions and notation.}
\end{table}

\section{Divergence--type conformal theories}
\label{sec2}

In this section we give a brief review about divergence-- type fluid theories within the framework of General Relativity. We follow closely the work by Geroch and Lindblom \cite{geroch1990dissipative},  also referring
to~\cite{geroch1991causal,geroch1995relativistic,geroch2001hyperbolic,0264-9381-15-3-015,rezzolla2013relativistic}
where needed. We focus our analysis in the conformal case, for which we introduce further
concepts and notation that simplify arriving at our main results.

One of the simplest but physically consistent theories of dissipative relativistic fluids that may have a well posed initial value formulation are those of divergence type, in which dynamical equations can be written as total divergence equations. The simplicity of those theories lies
in the fact that they can be constructed from a single generating function (that is, a sufficiently smooth scalar field) and a dissipation--source tensor, both depending on the corresponding fluid variables. This sort of theories facilitates the understanding on how to describe dissipative fluids in the framework of General Relativity.

Let us consider a $d$-dimensional time-oriented background spacetime $(\mathcal{M}, g_{ab})$. The starting point is the assumption that any fluid theory is made up by the following characteristics \cite{geroch1990dissipative}:

\begin{itemize}

\item[\textit{(i)}] The dynamical variables of the theory are the energy--momentum tensor $T^{ab}$ [a (2,0) symmetric tensor field], and a four--vector field $N^a$ representing the particle number current density of the flow;

\item[\em (ii)] The evolution of these variables is governed by the set of first order partial differential equations:
\begin{eqnarray}\label{formalism}
\nabla_a N^a &=& 0 \label{pcons}\\
\nabla_a T^{ab} &=& 0 \label{Tabcons} \\
\nabla_a A^{abc} &=& I^{bc}. \label{dissipeq} 
\end{eqnarray}
Here, the $(3, 0)$ tensor field $A^{abc}$ defined over $\mathcal{M}$ is symmetric and trace--free in the last two indices, and it is an algebraic function on the fluid variables. The tensor $I^{ab}$ is also symmetric and trace-free, and depends algebraically on the fluid variables as well.

\item[\em (iii)] There exists a vector field $S^a$  --also a local, algebraic function of the fluid variables--, and as a consequence of equations (\ref{formalism}), (\ref{Tabcons}) and (\ref{dissipeq}), it satisfies the following inequality:
\begin{equation}\label{entropycons}
\nabla_a S^a \geq 0.
\end{equation}
\end{itemize}

The first two equations (\ref{pcons}) and (\ref{Tabcons}) are respectively the familiar particle number and energy--momentum conservation laws for relativistic fluids. The third equation furnishes a description of the dissipative properties of the fluid, and provides ``constitutive relations'' for the theory. Notice that the symmetries of the dissipative tensor $A^{abc}$ implies that the number of equations equals the number of variables.
Last, the inequality (\ref{entropycons}) suggests that $S^a$ has the meaning of \textit{entropy density} of the fluid. In fact, by integrating both sides of inequality (\ref{entropycons}) over the volume $V(\Sigma,\Sigma')$ limited by the space-like hypersurfaces $\Sigma,\;\Sigma'\subset\mathcal{M}$ with $\Sigma'$ in the future of $\Sigma$ and applying Stokes theorem one gets
\begin{eqnarray}
0 &\leq& \int_{V}{(\nabla_a S^a)\;\sqrt{-g}\;d^4x}\nonumber \\
&=& \int_{\Sigma'}{S^a\; d\Sigma'_a} - \int_{\Sigma}{S^a\; d\Sigma_a}, \nonumber
\end{eqnarray}
from which the quantity
\begin{equation}
\mathcal{S}^a(\Sigma) := \int_{\Sigma}{S^a\; d\Sigma_a}
\end{equation}
is non--decreasing.

Throughout this work, we shall concentrate on \textit{conformally invariant} fluids, for which there is no particle number conservation. From a thermodynamical point of view, one may think that, as in the case of a photon gas, the chemical potential is zero, since the process of creating photons does not cost energy (see \cite{doi:10.1119/1.1336839} for a detailed discussion). In fact, one should take the limit of the chemical potential tending to zero in order to connect a particle gas to a photon gas. On the other hand, recall that in the ultra-relativistic limit, particle rest energy density becomes irrelevant, and particles move at an energy scale much higher that the one they have at rest (thus, getting closer to photon dynamics). Henceforth, as custommary in
this regime, we shall discard the conservation equation (\ref{pcons}) and consider just (\ref{Tabcons}) and (\ref{dissipeq}) as dynamical equations of the fluid.

A key observation within this framework is the following: condition \textit{(iii)} does not hold for all $N^a$ and $T^{ab}$, but only for those that represent a thermodynamic process, that is, for those that satisfy the conservation equations. Thus, requiring the existence of such an entropy law, together with
the symmetry of the energy--momentum tensor, implies the existence of \textit{new variables}
$\{\xi_a, \xi_{ab}\}$\footnote{When particle conservation is included, it is normal to consider $\xi_{ab}$ to be \textit{trace--free}, 
for this is the freedom remaining in the energy-- momentum tensor. Alternatively, one can think that the trace scalar freedom can be 
taken care with a scalar variable arising in this formalism from particle number conservation (that is, the associated Lagrange multiplier). Here we do not require it, 
but nonetheless we shall see it will appear as a requirement.} and a \textit{generating function}, $\chi(\xi_a, \xi_{ab})$ such that, 
\begin{eqnarray}
T^{ab} &\equiv& \frac{\partial^{2}\chi}{\partial \xi_a \partial \xi_b} \label{Tfromchi} \, , \\
A^{abc}&\equiv& \frac{\partial^{2} \chi}{\partial \xi_{a} \partial \xi_{bc}}  \label{Afromchi} \, .
\end{eqnarray}
The existence of these variables come out as Lagrange multipliers of the equations of motion \cite{geroch1990dissipative,pennisi87}. Therefore, a single scalar function of the variables $\chi(\xi_a, \xi_{ab})$ suffices to formally describe
the behavior of the fluid. We shall see that the tensor field $\xi_{ab}$ is associated with (and actually encodes) departures from a perfect fluid description. This is because when considering only perfect fluids, the dissipative equation (\ref{dissipeq}) does not appear, so there is no associated Lagrange multiplier, namely $\xi_{ab}$. It is natural then to consider the generating function $\chi$ as an expansion in terms of the dissipative scalar--type variables
(i.e. scalar fields defined in terms of $\xi_{ab}$). 

The most general function (up to second order in dissipative variables) that may be constructed in this scheme can be expressed as
\begin{equation}
\chi(\mu,\nu,\psi) = \chi^{0}(\mu) + \chi^{1}(\mu)\nu + \sum_{i=0}^{3}{\chi^{2}_i (\mu)\psi^i}\;, \label{chiexpansion}
\end{equation}
where $\mu := \xi^{c}\xi_{c}$ is the square of the norm of $\xi^c$, $\nu := \xi^{ab}\xi_{a}\xi_{b}$, and the second order scalars
\begin{equation}
  \psi_{1} := \xi^{ab}\xi_{ab}, \qquad \psi_{2}:= \xi^{ab}\xi_{b}\xi_{ac} \xi^{c}, \qquad \psi_{3}:=\nu^{2}. 
\end{equation}
The entropy current in this framework is determined by the generating function $\chi$ as \cite{geroch1990dissipative}
\begin{equation}\label{eq:S_def}
S^a = \frac{\partial \chi}{\partial \xi_a} - \xi_b T^{ab} - \xi_{bc} A^{abc}\,,
\end{equation}
and it satisfies
\begin{equation}
\nabla_a S^a = -\xi_{ab}I^{ab} \label{eq:divS_def}  \, .
\end{equation}
Consequently, entropy production in this framework is governed by the 
divergence of $A^{abc}$ and $\xi_{ab}$. The latter also justifies our prior statement that $\xi_{ab}$
can be regarded as an intrinsically ``dissipative'' variable.
In what follows, we will work order by order and construct the generating function $\chi$ as a linear combination of those order--contributions. We shall see that
the requirement of \textit{conformal invariance} has a significant effect in determining the possible different
contributions. Before we present our findings, we discuss general consequences of conformal symmetry 
which will be useful throughout this work.

\section{Conformal Invariance}
\label{sec3}

In this section we review conformal transformations, and derive some algebraic properties for the dynamical variables that guarantee conformal invariance of the evolution equations. After a discussion of the general structure of some conformally invariant tensor fields we shall make use of, we introduce the notion of \textit{conformal weights}, and scaling of some quantities of interest with the conformal factor.

\noindent Recall that under a conformal transformation
\begin{equation}\label{conftransf}
\hat{g}_{ab} = \Omega^{2} g_{ab},
\end{equation}
the connection $\nabla_c$ changes as \cite{wald2010general}
\begin{equation}
\hat{\nabla}_{a}X^{b} = \nabla_{a}X^{b} + C^{b}{}_{ac} X^{c},
\end{equation}
when acting over a vector field $X^c$. Here,
\begin{equation}
C^{a}{}_{bc} = g^{ad}\left(2\hat{n}_{(b} g_{c)d} - \hat{n}_{d} g_{bc}\right),
\end{equation}
with
\begin{equation}
\hat{n}_{c} := \frac{1}{\Omega}\nabla_{c} \Omega,
\end{equation}
over the points of $\mathcal{M}$ in which $\Omega\neq 0$. We recall as well the identity $C^{a}{}_{ac} = d\hat{n}_{c}$.

\textit{Conformal invariance} of the fluid equations imply the existence of two constants $\alpha$, $\beta$ such that they remain unaltered after a conformal transformation, i.e.,
\begin{equation}
\hat{T}^{ab} = \Omega^{\alpha}\; T^{ab}\;; \qquad \hat{A}^{abc} = \Omega^{\beta}\; A^{abc}
\end{equation}
with
\begin{equation}
\hat{\nabla}_a \hat{T}^{ab} = 0\;; \qquad \hat{\nabla}_a\hat{A}^{abc} = \hat{I}^{bc}.
\end{equation}
Thus, for the energy--momentum tensor, a conformal transformation implies 
\begin{eqnarray*}
\hat{\nabla}_{a}\hat{T}^{ab} &=& \nabla_a \hat{T}^{ab} + C^a{}_{ac} \hat{T}^{cb} + C^b{}_{ad}\hat{T}^{ad} \\
&=& \Omega^{\alpha}\left[\nabla_a T^{ab} + \left(\alpha + d + 2\right) \hat{n}_a T^{ab} - \hat{n}^b T^{ac}g_{ac}\right]
\end{eqnarray*}
for all $\hat{n}_a$. Conformal invariance implies the last two terms of the right hand side must vanish, i.e.,
\begin{equation}
\left(\alpha + d + 2\right) \hat{n}_a T^{ab} - \hat{n}^b T^{ac}g_{ac} = 0.
\end{equation}
Contracting with $\hat{n}_b$ and requiring the above equation holds for any $\hat{n}_a$ we get that both terms must cancel separately, yielding
\begin{equation}
\alpha = - (d+2), \qquad g_{ab}T^{ab} = 0.
\end{equation}
Therefore, any trace-free energy momentum tensor transforming like
\begin{equation}
\hat{T}^{ab} = \Omega^{-(d+2)}\; T^{ab}
\end{equation}
under conformal transformations will have a conformal invariant conservation law.

Let us turn now our attention to the constitutive relation tensor, $A^{abc}$. A conformal transformation for its divergence implies,  
\begin{eqnarray*}
\hat{\nabla}_{a}\hat{A}^{abc} &=& \nabla_{a}\hat{A}^{abc}  + C^{a}{}_{ad} \hat{A}^{dbc} + 2C^{(b}{}_{ad} \hat{A}^{|a|c)d} \\
&=& \nabla_{a}\hat{A}^{abc}  + d\hat{n}_{a} \hat{A}^{abc} + g^{db}(2\hat{n}_{(a} g_{m)d} - \hat{n}_{d} g_{am})\hat{A}^{acm} \nonumber \\ 
&& + g^{dc}(2\hat{n}_{(a} g_{m)d} - \hat{n}_{d} g_{am})\hat{A}^{abm} \\
&=& \nabla_{a}\hat{A}^{abc}  + d\hat{n}_{a} \hat{A}^{abc} + g^{db}(\hat{n}_{a} g_{md} + \hat{n}_{m} g_{ad} - \hat{n}_{d} g_{am})\hat{A}^{acm} \\  
&& + g^{dc}(\hat{n}_{a} g_{md} + \hat{n}_{m} g_{ad} - \hat{n}_{d} g_{am})\hat{A}^{abm} \\
&=& \nabla_{a}\hat{A}^{abc}  + d\hat{n}_{a} \hat{A}^{abc} 
+ \hat{n}_{a}\hat{A}^{acb}  + \hat{n}_{d}\hat{A}^{bcd}  - \hat{n}^{b} g_{ad}\hat{A}^{acd}  
+ \hat{n}_{a}\hat{A}^{abc} \\  
&& + \hat{n}_{d}\hat{A}^{cbd}  - \hat{n}^{c} g_{ad}\hat{A}^{abd} \\
&=& \nabla_{a}\hat{A}^{abc}  + (d+2)\hat{n}_{a} \hat{A}^{abc} 
  + 2\hat{n}_{d}\hat{A}^{(bc)d}  - 2\hat{n}^{(b} g_{ad}\hat{A}^{|a|c)d}\\
&=& \Omega^{\beta}\left[\nabla_{a} A^{abc}  + (\beta + d + 2)\hat{n}_{a}        A^{abc} 
  + 2\hat{n}_{d} A^{(bc)d}  - 2\hat{n}^{(b} g_{ad} A^{|a|c)d}\right]
\end{eqnarray*}

As with the previous case, the second, third and fourth terms of the right hand side must vanish, i.e., 
for arbitrary $\hat{n}_{a}$,
\begin{equation}
(\beta + d + 2)\hat{n}_{a}        A^{abc} 
  + 2\hat{n}_{d} A^{(bc)d}  - 2\hat{n}^{(b} g_{ad} A^{|a|c)d} = 0. \label{Aconformal}
\end{equation}
Contracting (\ref{Aconformal}) with $\hat{n}_{b} \hat{n}_{c} $ gives
\begin{equation}
\frac{1}{2}(\beta + d + 4)A^{abc}\hat{n}_{a}\hat{n}_{b} \hat{n}_{c}  - A^{bca}g_{bc} \hat{n}_a = 0
\end{equation}
for all $\hat{n}_a$. Since $A^{abc}$ is trace free in the last two indices, each term must vanish separately.
Thus, $\beta = -(d+4)$, and $A^{bca}g_{bc} =0$. Therefore, equation (\ref{Aconformal}) reduces to
\begin{equation}\label{Acondition}
 \hat{n}_a A^{abc} -  \hat{n}_{a}A^{(bc)a} =0,  
\end{equation}
for a totally trace--free tensor symmetric in the last two indices. Thus, $A^{abc}$ must transform as
\begin{equation}
\hat{A}^{abc} = \Omega^{-(d+4)} A^{abc}.
\end{equation}

As a conclusion, we may say that under a transformation like (\ref{conftransf}), conformal invariance 
is guaranteed if $T^{ab}$ is trace--free and $A^{abc}$ satisfies the relation (\ref{Acondition}).

\subsection{Conformal weights}
\label{CW's} 
Throughout this work, it will be important to
keep track of the ``conformal weights'' different quantities have. To this
end we define the \textit{conformal weight} $\mathcal{CW}$ of an arbitrary quantity $X$ as
\begin{equation}
\mathcal{CW}(X) := n \qquad \mbox{if} \qquad \hat{X} = \Omega^{-n} X,
\end{equation} 
where $\hat{X}$ is the conformally related quantity via the transformation (\ref{conftransf}). From our previous discussion we have that $\mathcal{CW}(T^{ab}) = d+2$, and $\mathcal{CW}(A^{abc}) = d+4$. 

On the other hand, for the entropy production equation (\ref{eq:divS_def}) to be \textit{conformally invariant} we must have $\mathcal{CW} (S^a) = d$. To see this, let $\gamma_1$ and $\gamma_2$ be scalars such that $\hat{S}^a = \Omega^{\gamma_1} S^a$ and $\hat{\sigma} = \Omega^{\gamma_2} \sigma$, where $\sigma = -\xi_{ab} I^{ab}$ (see equation (\ref{eq:divS_def})). By similar arguments to the ones given before, we get
\begin{equation}\label{entropy-trans}
\hat{\nabla}_a\hat{S}^a = \Omega^{\gamma} \left(\nabla_a S^a + (\gamma + d) \hat{n}_a S^a\right).
\end{equation}
for all $\hat{n}_a$. Requiring the left hand side of (\ref{entropy-trans}) to be equal to $\hat{\sigma}$, we must have $\gamma_1 = \gamma_2 = -d$, from which we conclude that $\mathcal{CW}(S^a) = d$. Now, it is possible to use our definition of $S^a$ (\ref{eq:S_def}) to assert that, by virtue of the conformal weights of $T^{ab}$ and $A^{abc}$ computed before, $\mathcal{CW}  (\xi_a) = -2$ and $\mathcal{CW}(\xi_{ab}) = -4$. This, in turn, implies that $\mathcal{CW}(\xi^a) = \mathcal{CW}(\xi^{ab}) = 0$.

Next, from the conformal weights of $\xi^a$ and $\xi_a$ we get $\mathcal{CW}(\mu) = -2$. Likewise, notice also that as a consequence of (\ref{eq:S_def}),
\[
d = \mathcal{CW}\left(\frac{\partial \chi}{\partial \xi_a}\right) = \mathcal{CW}(\chi) - \mathcal{CW}(\xi_a),
\]
from which $\mathcal{CW}(\chi) = d - 2$.

These relations, in particular, will help to uniquely determine the powers of $\mu$ in the different factors that will appear order by order through our analysis. In what follows, we derive expressions for the quantities $T^{ab}, A^{abc}$ and constitutive relations order by order with respect to dissipative contributions [recall equation (\ref{chiexpansion})].

\section{Perfect fluid and equilibrium states}
\label{perfect-fluids}

In this section we give a detailed description of the theories of conformal fluids without dissipation. We will see that from these equations emerges directly the familiar perfect fluid structure of the energy--momentum tensor, with radiation equation of state. Next, following the guidelines of~\cite{geroch1990dissipative}, we shall give a first characterization of the equilibrium states of these theories, as well as we compute the entropy as a function of the fundamental variables, verifying that it is a conserved quantity in equilibrium.

We begin with the zeroth order in the expansion (\ref{chiexpansion}).
To this order we get, from equation (\ref{Tfromchi}), 

\begin{equation}\label{Tabzeroth}
T_o^{ab} = 4\chi^o_{\mu\mu} \xi^a \xi^b + 2\chi^o_{\mu} g^{ab},
\end{equation}
where the subindex $\mu$ denotes derivative with respect to $\mu$, and from equation (\ref{Afromchi}),
\begin{equation}
A_o^{abc} = 0.
\end{equation}
Next, requiring the energy momentum tensor to be trace--free, we obtain the following
condition for $\chi^o$: 
\begin{equation}\label{traceT0}
2d\chi^{o}_{\mu} + 4\mu \chi^{o}_{\mu\mu} = 0, 
\end{equation}
The physically valid solution to this equation is
%
%
%
%
%


\begin{equation}\label{chisol-zeroth}
\chi^{o}(\mu) = \frac{\chi^{o}_{o}}{\mu^{\frac{d}{2}-1}}\;,
\end{equation}
up to an irrelevant constant $\chi^{o}_{o}$ which, we shall see, does not contribute to the equations.

The solution thus obtained corresponds to a perfect fluid. To see this, let us assume that $\xi^a$ is time--like and introduce the vector field
\[
u^a := \frac{\xi^a}{\sqrt{-\mu}}.
\]
Next, by introducing thermodynamical variables $\{\rho, p\}$ generally defined as
\begin{equation} \label{ro-p-defs}
\rho := T^{ab} u_a u_b\;, \qquad p := \frac{T^{ab}\left(g_{ab} + u_a u_b\right)}{d-1}\;,
\end{equation}
it is straightforward that the expression (\ref{Tabzeroth}) is equivalent to
\begin{equation}
T_{PF}^{ab} = (\rho + p)u^{a}u^{b} + p g^{ab},
\end{equation}
via the identification
\begin{equation}
p := 2\chi^{o}_{\mu}\;,\qquad \rho := -4 \mu \chi^{o}_{\mu\mu} - 2 \chi^{o}_{\mu}.
\end{equation}

At this point, several observations should be pointed out.
\begin{itemize}

\item In order to make an identification like the one made above, it is not strictly necessary to assume that $\xi^a$ is time-like. Nevertheless, if the aim is to describe a perfect fluid with 4--velocity $u^a$, a natural requirement is that it be.

\item As expected, the stress-energy tensor we have already derived corresponds to a pure radiation perfect fluid, so the conformal invariance gives to this order a unique perfect fluid. In fact, a direct consequence of the trace--free equation (\ref{traceT0}) yields the corresponding state equation, namely
\begin{equation}
p = \frac{\rho}{d-1},
\end{equation}
as it can be straightforwardly checked through (\ref{ro-p-defs}).

\item The positivity of the conformal density $\rho$ is a consequence of the energy conditions demanded from $T_o^{ab}$. In fact, it is straightforward to check that if $T_o^{ab}$ satisfies the dominant energy condition, and $\xi^a$ is time-like, then $\rho \geq 0$.

\item For each $x\in\mathcal{M}$, the vector $\xi^a$ is an eigenvector of $T_o^{a}{}_b$. Thus, if $\xi^a$ is assumed to be time--like, it spans the 1--dimensional time--like eigenspace of $T_o^{a}{}_b$, and $u^a$ is the so--called ``Landau frame" (see, for instance, \cite{rezzolla2013relativistic}).

\item Introducing the quantity
\begin{equation}\label{temperature-def}
T := \frac{1}{\sqrt{-\mu}},
\end{equation}
so that $u^a = T \xi^a$, and using the explicit solution (\ref{chisol-zeroth}) for the generating function, we see that both energy density and pressure scale as $T^d$. Also, it is straightforward that $\mathcal{CW}(T) = 1$ (see section \ref{CW's} above). We shall identify this quantity with the \textit{temperature} of the fluid. In this context, the notion of temperature captured in (\ref{temperature-def}) has been extensively studied in the past, in order to characterize thermal equilibrium states of gravitating systems. In particular, we highlight Tolman's work \cite{Tolman:1930ona}, wherein a expression like (\ref{temperature-def}) is derived for static spacetimes.
\end{itemize}

We can also derive further useful relations at this order.
The conservation equation, projected along the fluid's velocity, $u_a \nabla_b T^{ab} = 0 $, implies 

\begin{eqnarray}
\nabla_a u^a &=& -\frac{d-1}{d} u^a \nabla_a \ln p \nonumber \\
             &=& -(d-1) u^a\nabla_a \ln T \nonumber \\
             &=& \frac{d-1}{2} u^a\nabla_a \ln \left(- \mu\right) \, .
\end{eqnarray}

On the other hand, the projection on the orthogonal plane with respect to the fluid velocity yields 
\begin{eqnarray}
u^a \nabla_a u^b &=& -D^b \ln T \nonumber \\ 
&=& \frac{1}{2} D^b \ln \left(- \mu\right),
\end{eqnarray}
where $D_a := \left(\delta_a{}^b + u_a u^b\right) \nabla_b$ is the derivative of the hypersurface orthogonal to the integral lines of $u^a$.

It is useful to also obtain related expressions in terms of $\xi^a$. In fact, taking the divergence of the full expression (\ref{Tabzeroth}), namely

\begin{eqnarray}\label{divT0}
\nabla_a T_o^{ab}  &=& 2\chi^o_{\mu\mu} \left (g^{ab} -\frac{d}{\mu} \xi^a \xi^b \right) \nabla_a \mu - \frac{2d}{\mu}\chi_{\mu} \left ( D\xi^b + \dot{\xi}^b - \frac{\dot{\mu}}{\mu}\xi^b \right) \nonumber \\
&=& \chi^o_{\mu}\frac{d}{\mu}\left [\left(\frac{d + 2}{\mu} \dot{\mu} - 2D\right) \xi^b -  \nabla^b \mu - 2\dot{\xi}^b \right]\;,
\end{eqnarray} 
where we have defined,
\begin{equation}
D := \nabla_a \xi^a\;,\quad \dot{\mu}:= \xi^a \nabla_a \mu\;,\quad \dot{\xi}^b := \xi^a \nabla_a \xi^b\;,
\end{equation}
from which it follows the identity 
\begin{equation}\label{xi-a}
\xi^b \dot{\xi}_b = \frac{1}{2}\dot{\mu}.
\end{equation} 
Now, contracting equation (\ref{divT0}) with $\xi_b$, one obtains
\begin{eqnarray}
0 &=& \xi_b \nabla_a T_o^{ab} \nonumber \\
&=& d \chi^o_{\mu}\left [\frac{\dot{\mu}}{\mu} d - 2D\right ]\;, \nonumber \end{eqnarray}
from which 
\begin{equation} \label{zeroth}
\frac{\dot{\mu}}{\mu} = \frac{2D}{d}.
\end{equation}
Last, replacing the above relation in the conservation equation (\ref{divT0}), one gets
\begin{eqnarray}
0 &=& \nabla_a T_o^{ab} \nonumber \\
&=& \chi^o_{\mu}\frac{2d}{\mu}\left [\frac{\dot{\mu}}{\mu}\xi^b - \frac{1}{2}\nabla^b \mu - \dot{\xi}^b \right]\;, 
\end{eqnarray}
which implies the identity
\begin{equation}
\dot{\xi}^b = \frac{\dot{\mu}}{\mu}\xi^b - \frac{1}{2}\nabla^b \mu.
\end{equation}

\subsection{Entropy density contribution}
Let us compute the entropy current at this order. Following the definition given in equation (\ref{eq:S_def}), we get
\begin{eqnarray}
S_o^a &=& \frac{\partial \chi^o}{\partial \xi_a} - \xi_bT_o^{ab} \nonumber \\
&=& - 4 \mu \chi^o_{\mu \mu}\xi^a \nonumber \\
&=& \frac{d}{d-1}\rho \xi^a \\
&=& \frac{d}{d-1}\frac{\rho}{T} u^a.
\end{eqnarray}
For the divergence we get
\begin{eqnarray}
\nabla_a S_o^a &=& 2dD\chi^o_{\mu} + 2d\xi^a\nabla_a\chi^o_{\mu}\nonumber \\
&=& 2dD\chi^o_{\mu} + 2d\chi^o_{\mu\mu}\dot{\mu} \nonumber \\
&=& 2d\chi^o_{\mu}\left(D-\frac{d\dot{\mu}}{2\mu}\right) = 0,
\end{eqnarray}
by virtue of relation (\ref{zeroth}). Thus, there is no change of entropy to this non--dissipative
order, as expected.

\subsection{Equilibrium states}

In this section we discuss the equilibrium states of this theory. In line with the
arguments of Geroch and Lindblom~\cite{geroch1990dissipative,geroch1991causal}, these states must have the same properties as those of Eckart theory; that is, a rigid flow and a constant and stable temperature field. They refer to equilibrium states those solution of fluid equations in which dynamics is time reversible. With simple algebraic arguments, the authors prove that those states have the following properties: the source tensor $I^{ab}$ evaluated on equilibrium states is zero; there is no entropy production and the vector field $\xi^a$ is a Killing vector field.

We now give a first argument to assert that equilibrium states for conformal dissipative relativistic theories are those such that $\xi^a$ is a conformal Killing vector field. Recall that, in general, $X^a$ is a \textit{conformal Killing vector field} if there exists a scalar field $\alpha$ such that \cite{wald2010general}
\[
\nabla_{(a}X_{b)} = \alpha g_{ab}.
\]
If this relation holds for some $\alpha$, contraction with $g^{ab}$ over both sides of it implies that $\alpha = \nabla_a X^a / d$.

To see the nature of $\xi^a$, let us compute the symmetrized derivative of it, which is a symmetric tensor of rank $(0,2)$. The unique symmetric tensor fields that can be constructed in terms of $\xi^a$ and the spacetime metric are the metric itself, and $\xi_a\xi_b$. So it is natural to propose that
\begin{equation}\label{ansatz-killing}
\nabla_{(a}\xi_{b)} = t\; g_{ab} + s\;\xi^a\xi^b,
\end{equation} 
with $t,s$ functions of $\mu$ and the dimension, to be determined. Contraction of (\ref{ansatz-killing}) with $g^{ab}$ gives
\[
D = td + s\mu,
\]
while its contraction with $\xi^a\xi^b$ yields now the relation
\[
\frac{\dot{\mu}}{2} = t\mu + s\mu^2.
\]
Recalling the identities (\ref{xi-a}) and (\ref{zeroth}), the above two relations imply $t = D/d$ and $s = 0$, so $\xi^a$ is a Killing vector field. Thus, equilibrium states in these theories are those such that $\xi_{ab} = 0$ (that trivially holds in this case) and $\xi^a$ is a conformal Killing vector field.

\section{First order theory}
\label{1storder-section}

In this section we give a description of the first order contribution, which stems from adding linear dissipative--like terms to the dynamical variables. This is a better approximation for describing fluids when microscopic time-scales are comparable with the macroscopic ones, since the local thermodynamic equilibrium breaks down (e.g.~\cite{rezzolla2013relativistic}). At this order, we shall find the four--velocity--orthogonal heat flux, deriving  a Fourier--like equation, as well as the corresponding shear viscosity to the stress-energy tensor. We next compute the entropy current and include a discussion about the Landau frame at this order, from which we shall see that energy dissipation and, thus, entropy production, plays a central role. 

Following the orthonormal decomposition discussed in Appendix~\ref{app1}, it is convenient to decompose the dissipative variable $\xi_{ab}$ as,
\begin{equation}
\xi_{ab} = \frac{\nu}{\mu^{2}}\xi_{a}\xi_{b} + \frac{2}{\mu}\xi_{(a} r_{b)} + \tau_{ab}\;, \label{decomposixiab}
\end{equation}
where we have introduced the quantities
\begin{eqnarray}
r_{a} &\equiv& \xi_{ab}\xi^{b} - \frac{\nu}{\mu}\xi_{a} \,, \label{qdef} \\
\tau_{ab} &\equiv& \tilde{\tau}_{ab} - \frac{h_{ab}}{d-1}\frac{\nu}{\mu}\;. \label{taudef}
\end{eqnarray}
with the straightforward conditions
\begin{equation} \label{tauperpxi}
\tau_{ab}\xi^{b} = 0\;, \qquad r_{a}\xi^{a} = g^{ab} \tilde \tau_{ab} = 0.
\end{equation}

Notice that the unique vector fields that can be constructed as algebraic functions of the variables $(\xi_a, \xi_{ab})$ and the metric $g_{ab}$, which are (at most) linear in $\xi_{ab}$ are $\xi^a := g^{ab}\xi_b$ and $\ell^a:=\xi^{ab}\xi_a$. In particular, if $\xi^a$ is time--like, then $r^a$ is always spatial and orthogonal to $\xi^a$.
We shall make use of this decomposition below to uncover the constitutive
relations at this order. 

The contribution to the generating function at first order is
\[
\chi^{(1)}(\mu,\nu) = \chi^1(\mu)\nu\;,
\]
where $\chi^1$ is an arbitrary function of $\mu$, and $\nu:=\xi^{ab}\xi_a\xi_b$ is the unique (non trivial) scalar which is linear in $\xi_{ab}$ and is an algebraic function of the variables $(\xi_a, \xi_{ab})$ and the metric. The corresponding contributions of the stress energy tensor and the constitutive tensor that arise from $\chi^{(1)}(\mu,\nu)$ are, respectively,
\begin{equation}\label{T11tensor}
T^{1}_{ab} = 4 \chi^{1}_{\mu\mu} \nu \xi_{a} \xi_{b} + 8\chi^{1}_{\mu} \xi_{(a} \xi_{b)c} \xi^{c} + 2 \chi^{1}_{\mu} \nu g_{ab} + 2\chi^{1}\xi_{ab}\, ,
\end{equation}
\begin{equation} \label{A1tensor}
A_1^{abc} = \chi^{1}(2g^{a(b}\xi^{c)} - \frac{2}{d}\xi^{a}g^{bc}) + 2\chi^{1}_{\mu}\xi^{a} (\xi^{b}\xi^{c} - \frac{\mu}{d}g^{bc}).
\end{equation}
Now, let us impose the conformal invariance requierements to those expressions. Trace--free condition for (\ref{T11tensor}) implies the following two equations,
\begin{eqnarray}
2\mu \chi^{1}_{\mu\mu} + (d+4) \chi^{1}_{\mu}  &=& 0 \, ,\label{eq:focd} \label{der-1st-rd}\\
g^{ab}\xi_{ab} &=& 0 \, .
\end{eqnarray}
The second one was expected, since $\xi_{ab}$ is trace-free by construction. The former is satisfied if and only if
\begin{equation} \label{chi1soln}
\chi^{1} = \chi^1_{1} + \frac{\chi^1_o}{\mu^{\frac{d+2}{2}}} \, ,
\end{equation}
with real constants $\chi^1_{1}$ and $\chi^1_o$. 

Imposing now the corresponding requirement for (\ref{A1tensor}) [given by relation (\ref{Acondition})], one gets
\begin{equation} \label{eqchi1storder}
2\mu \chi^1_{\mu} + (d+2) \chi^1 =0,
\end{equation}
which eliminates the constant $\chi^1_{1}$ in the solution (\ref{chi1soln}). It is crucial here to notice that relations (\ref{eqchi1storder}) and (\ref{der-1st-rd}) are compatible with each other, in the sense that one is a first integral of the other.
Therefore, equation (\ref{eqchi1storder}) allows us to express all the coefficients for $T^1_{ab}$ and $A_1^{abc}$ in terms of a \textit{single} function, $\chi^1$. Indeed, we get
\begin{equation}
T_1^{ab} = \chi^1\left[\frac{(d+2)(d+4)}{\mu^2} \nu \xi^a \xi^b - \frac{4(d+2)}{\mu} \xi^{(a} \xi^{b)c}\xi_c - \frac{d+2}{\mu} \nu g^{ab} + 2 \xi^{ab}\right].
\end{equation}
and
\begin{eqnarray} \label{A1final}
A_1^{abc}& = & \chi^1\left[2g^{a(b} \xi^{c)} + \xi^a g^{bc} - \frac{d+2}{\mu} \xi^a\xi^b\xi^c\right] \nonumber \\
                & = & \chi^1\left[2h^{a(b} \xi^{c)} + \xi^a\left(g^{bc} - \frac{d}{\mu}\xi^b \xi^c\right)\right] \nonumber \\
                & = & {\chi^1}{\sqrt{-\mu}}\left[2h^{a(b} u^{c)} + u^a h^{bc} + (d-1)u^a u^b u^c)\right] \,,
\end{eqnarray}
where we have introduced, one more time, the normalized time--like vector field
\begin{equation} \label{u}
u^c = \frac{\xi^c}{\sqrt{-\mu}},
\end{equation}
in the last line of the above calculation. Notice that $A_1^{abc}$ is trace--free in all entries (recall discussion around equation (\ref{Acondition})), and it can be 
expressed in the form given in equation (\ref{Atensor})  of Appendix A by choosing $V^a = \xi^a$ and
\begin{equation} \label{Sab-1storder}
S^{ab} = g^{ab} - \frac{d+2}{3\mu} \xi^a \xi^b.
\end{equation}
Also, by imposing (\ref{Acondition}) to (\ref{A1tensor}) we obtain again the first integral (\ref{eqchi1storder}).

On the other hand, we can compute explicitly the divergence of $A_1^{abc}$, that will be used later on. Indeed, we get
\begin{eqnarray}
\nabla_a A_1^{abc} \label{diva1storder}
= &-&\frac{\chi^1(d+1)}{2\mu}\left[2\sqrt{-\mu}\;u^{(b}D^{c)}\mu + \dot{\mu} h^{bc} + (d-1)\dot{\mu}u^b u^c \right]\nonumber \\
&+& \chi^1\sqrt{-\mu}\left[2D^{(b}u^{c)} + (\nabla_a u^a) h^{bc}\right. \nonumber \\
&+& \left.(d+1)\left((\nabla_a u^a) u^b u^c + 2u^a u^{(b}\nabla_a u^{c)}\right)\right] \,.
\end{eqnarray}

To find out physical quantities and constitutive fluid relations, let us now use the decomposition (\ref{decomposixiab}) to re-express $T^1_{ab}$ and $A_1^{abc}$ in order to discuss some consequences. We begin with $T^{1}_{ab}$, for which we get
\begin{equation} \label{T1posta}
T^{1}_{ab} = \chi^1\frac{\nu}{\mu^2}\frac{d^2(d+1)}{d-1}\left[\xi_a \xi_b - \frac{\mu}{d} g_{ab}\right] - \frac{4(d+1)\chi^1}{\mu} \xi_{(a}r_{b)} + 2 \chi^1 \tilde{\tau}_{ab}.
\end{equation}
Recalling relation (\ref{u}), one can easily deduce that the first order ``corrections'' to density and pressure are,
\begin{equation}\label{rho-p-firstorder}
\rho_1 = - d (d+1) \chi^1 \frac{\nu}{\mu}, \qquad  p_1 =  - \chi^1 \frac{d (d+1)}{d-1}\frac{\nu}{\mu} \, ,
\end{equation}
where the definitions (\ref{ro-p-defs}) have been used and, as expected, there is no change in the resulting equation of state, i.e.,
\begin{equation}\label{eqofstate}
p = \frac{\rho}{d-1},
\end{equation}
with
\[
\rho = \rho_0 + \rho_1, \quad p = p_0 + p_1,
\]
where $\rho_0$ and $p_0$ are the zeroth order contributions found earlier. Moreover, from the dominant energy condition for the full $T^{ab}$, $\rho > 0$.

\subsection{Heat flux and a Fourier-like equation}

In non relativistic frameworks, one usually introduces the notion of heat flux in the corresponding co--moving frame as a spatial vector $\vec{q}$ that measures the rate of energy flow per unit area at each point of the space. The corresponding heat equation is a consequence of the first and second law of Thermodynamics, plus the well known \textit{Fourier equation} \cite{landau2013fluid}
\begin{equation}\label{q-no-cov}
\vec{q} = - k\; \vec{\nabla}T,
\end{equation}
with $k$ the \textit{thermal diffusivity} and $T$ the temperature. As discussed in the introduction, a generalization of these ideas within a relativistic framework is not straightforward for several reasons \cite{Smerlak2012}. In particular, equation (\ref{q-no-cov}) leads to the following parabolic equation for $T$:
\begin{equation}
\partial_t T = a\; \nabla^2 T,
\end{equation}
where $a$ is a positive constant and $\nabla^2$ is the Laplace operator. While this equation possesses a well posed initial value problem (see a discussion in the textbook \cite{kreiss2004initial}), its nature provides no bound for the speed of propagation of thermal disturbances\footnote{One can easily see that this is already the situation by solving the associated one--dimensional initial value problem over a finite domain, with an initial profile of temperature given by a localized heat impulse.}, thus making difficult its identification in the relativistic regime.

Nonetheless, given a theory of relativistic fluids with energy--momentum tensor $T^{ab}$, it is possible introduce
the notion of heat flux in the co--moving frame $u^a$, as the projection of the energy flux onto the hypersurface that is orthogonal to $u^a$, namely
\[
q^a := - h^{ab}T_{bc}u^c,
\]
which satisfies a constitutive relation of the form (see, for instance, \cite{landau2013fluid})
\begin{equation}\label{sim}
q^a \sim - h^{ab}\left(\nabla_b T + T a_b\right),
\end{equation}
where $a^c = u^b\nabla_b u^c$, and $h^{ab}$ the induced metric to the hypersurface orthogonal to $u^c$, namely 
\[
h^{ab} = g^{ab} + u^a u^b.
\]

In accordance with expression (\ref{T1posta}), we identify the heat flux as
\begin{equation} \label{heatfluxfinal}
q^a := \frac{2(d+1)\chi^1}{\sqrt{-\mu}} r^a.
\end{equation}

We shall now derive a constitutive relation for (\ref{heatfluxfinal}) like the one in (\ref{sim}) as a direct consequence of the constitutive equation (\ref{dissipeq}) stated before. To this end, and recalling the orthonormal decomposition discussed in Appendix \ref{app1}, introduce the most general form for the tensor $I_{ab}$ that is trace-free and positive definite when contracting with a non-zero $\xi^{ab}$; namely,
\begin{equation} \label{source}
I_{ab} = \frac{\tau}{\gamma \mu} \xi_a \xi_b - \frac{2}{\kappa} \xi_{(a} r_{b)} - \frac{1}{\gamma} \tau_{ab} \, ,
\end{equation}
with $\{\kappa,\gamma\}$ positive functions of $\mu$ to be fixed and requiring $I_{ab}$ to have conformal weight $d$. We get
\begin{eqnarray} \label{entropyprod}
-\xi^{ab}I_{ab} &=& -\frac{\tau \nu}{\gamma \mu} + \frac{2}{\kappa}\ell^{a}r_a + \frac{1}{\gamma} \xi^{ab}\tau_{ab} \nonumber \\
&=& \frac{\tau^2}{\gamma} + \frac{2}{\kappa} r^a r_a + \frac{\tau^{ab}\tau_{ab}}{\gamma}\;, 
\end{eqnarray}
which is a positive function if $\kappa$ and $\gamma$ are positive, since $r^a$ is space--like, and $\tau^{ab}\tau_{ab} \geq 0$ (since $\tau_{ab}$ is a purely spatial tensor field). Also, we have used that the trace of $\tau_{ab}$ can be directly obtained using decomposition (\ref{decomposixiab}), namely $\tau = -\nu/\mu$.
Taking into account the decomposition performed in (\ref{source}), equation (\ref{dissipeq}) implies that
\begin{eqnarray} \label{heatflux}
r^d &=& - \frac{\kappa(\mu)}{\mu} \xi_c h^d{}_b \nabla_a A_1^{abc} \nonumber \\
&=& \chi^1 (d+1) \frac{\kappa(\mu)}{\mu} h^{db} \left(\frac{1}{2}\nabla_b \mu + \dot{\xi}_b\right) \nonumber \\
&=& \frac{\chi^1 (d+1)}{2} \frac{\kappa(\mu)}{\mu} \left(D^d \mu + 2h^{db}\dot{\xi}_b\right),
\end{eqnarray}

Now, notice that (\ref{heatflux}) can be expressed in a more suitable form. Indeed, the identity (\ref{divT0}) implies that
\begin{eqnarray} \label{proj}
h^{cb}\left(\frac{1}{2}\nabla_b \mu + \dot{\xi}_b\right) &=& - \frac{\mu}{2d\chi^o_{\mu}}h^c{}_{b}\nabla_a T_o^{ab} \nonumber \\
&=& -\frac{1}{\chi^o_o} \frac{\mu^{(d+2)/2}}{d(d-2)}h^c{}_{b}\nabla_a T_o^{ab},
\end{eqnarray}
where we have used the general solution for $\chi^o(\mu)$ obtained in the previous section (equation (\ref{chisol-zeroth}). Notice that relation (\ref{proj}) is \textit{off-shell}, i.e., it is a geometric identity that always holds, without assuming equations of motion. With this information, equation (\ref{heatflux}) reads
\begin{equation}\label{heatfluxFINAL}
r^a = - \frac{\chi^1_o}{\chi^o_o} \frac{d+1}{d(d-2)} \frac{\kappa(\mu)}{\mu} h^a{}_b \nabla_c T_o^{bc}\;.
\end{equation}

The above equation has a clear physical interpretation: the departure of these states from the equilibrium ones, carries away energy in form of ``heat flux". In particular, equilibrium states satisfy $r^a = 0$, but this is not the general case. 

Recalling expression (\ref{heatflux}) once more, it is possible to derive an analog of the Fourier law for first order dissipative fluids. In fact, by the identification indicated earlier,
$
T^2 \equiv -{\mu}^{-1},
$
we see that
\[
D^a T = h^{ab}\nabla_b T = \frac{\sqrt{-\mu}}{2\mu^2}D^a \mu,
\]
and equation (\ref{heatflux}) implies the vector field $r^a$ satisfies,
\[
r^a = - (d+1) \chi^1 \kappa \sqrt{-\mu} \left(D^a T + \frac{\sqrt{-\mu}}{\mu^2}h^{ab}\dot{\xi}_b\right),
\]
wich implies (via (\ref{heatfluxfinal})) the following equation for $q^a$:
\begin{equation} \label{fourierlaw}
q^a = - K \left(D^a T + \frac{\sqrt{-\mu}}{\mu^2}h^{ab}\dot{\xi}_b\right)\;,
\end{equation}
with
\begin{equation}
K:= 2 (d+1)^2 (\chi^1)^2 \kappa.
\end{equation}

Clearly, equation (\ref{fourierlaw}) suggests interpreting $q^a$ as a current energy density that is non-zero in general, and satisfies a Fourier--like transport equation (for references, see \cite{landau2013fluid,BALAZS1965222}). In particular, the variable transport coefficient $K$ is always positive, due to the positivity condition for the function $\kappa$ discussed in (\ref{entropyprod}).

Finally, a comment on the second term of the right hand side of equation (\ref{fourierlaw}) that could help us to further characterize the equilibrium states. Introducing the vector
\[
a^c := u^a\nabla_a u^c,
\]
and following the discussion of Landau and Lifshitz (chapter XV, section 126 of \cite{landau2013fluid}), we have that, in equilibrium, the projection of the energy--momentum conservation equation onto the hypersurface that is orthogonal to $u^a$ gives,
\[
(\rho + p) a_c = -\nabla_c p - u_c u^a \nabla_a p,
\]
from which
\begin{equation} \label{landau}
(\rho + p) h^{ac}a_c = - D^a p.
\end{equation}
On the other hand, a straightforward calculation gets
\[
h^{ac} \dot{\xi}_c = - \mu h^{ac} a_c,
\]
from which the Fourier Law (\ref{fourierlaw}) results,
\begin{equation} \label{fourierlawfinal}
q^a = - K \left(D^a T - \frac{T}{\rho + p} D^a p\right).
\end{equation}
Since in equilibrium we have $q^a = 0$, the above equation yields
\begin{equation}
D^a T - \frac{T}{\rho + p} D^a p = 0.
\end{equation}
in complete agreement with Landau's argument at the end of the chapter.

\subsection{Entropy density contribution}

According to this formalism, the first order contribution of the entropy current density $S^a$ is given by
\[
S_1^a = \frac{\partial \chi^{(1)}}{\partial \xi_a} - \xi_bT_1^{ab} - \xi_{bc}A_1^{abc},
\]
where $\chi^{(1)}(\mu,\nu)=\chi^1(\mu)\nu$. Explicitly, each term is given by
\begin{eqnarray*}
\frac{\partial \chi^{(1)}}{\partial \xi_a} &=& -(d+1)\chi^1\frac{\nu}{\mu}\xi^a + 2\chi^1r^a\;, \\
- \xi_b T_1^{ab} &=& -d(d+1)\chi^1\frac{\nu}{\mu}\xi^a + 2(d+1)\chi^1 r^a\;, \\
- \xi_{bc} A_1^{abc} &=& -2\chi^1 r^a + d\chi^1\frac{\nu}{\mu}\xi^a\;,
\end{eqnarray*}
so the first order contribution to the entropy is
\begin{equation}\label{entropy1st}
S_1^a = -\left ( d(d+1) + 1 \right )\chi^1\frac{\nu}{\mu}\xi^a + 2(d+1)\chi^1 r^a.
\end{equation}

The above expression agrees with standard results, \cite{hiscock1983stability}. In fact, by using our definition for heat flux presented in (\ref{heatfluxFINAL}), and recalling the definitions
\[
u^a = \frac{\xi^a}{\sqrt{-\mu}}, \qquad T = \frac{1}{\sqrt{-\mu}},
\]
as well as relations (\ref{rho-p-firstorder}) we get
\begin{equation}
S_1^a = s_1 u^a + \frac{q^a}{T},
\end{equation}
where the \textit{entropy density} $s_1$ up to this order is given by
\begin{equation}
s_1 = \frac{d(d+1)+1}{d(d+1)}\frac{\rho_1}{T}.
\end{equation}

Note that, as expected, $s_1\sim T^{d-1}$. From equation (\ref{eq:divS_def}) and expression (\ref{entropyprod}) the full entropy density satisfies
\begin{equation}
\nabla_a S^a = \frac{\tau^2}{\gamma} + \frac{2r^a r_a}{\kappa}  + \frac{\tau^{ab}\tau_{ab}}{\gamma},
\end{equation}
which implies the system creates entropy through shear and heat flux. This result is in agreement with the standard results (e.g.\cite{rezzolla2013relativistic}).

\subsection{Shear viscosity}

Returning now to expression (\ref{diva1storder}) and recalling equation (\ref{dissipeq}) as well as 
the decomposition (\ref{source}), we see that
\begin{equation}\label{tau}
\tau_{ab} = - \gamma \chi^1 \sqrt{-\mu} \left[2D_{(a}u_{b)} - (\nabla_c u^c) h_{ab}\right] + (d+1) \gamma \chi^1 \frac{\dot{\mu}}{2\mu} h_{ab}, 
\end{equation}
with trace
\begin{eqnarray} \label{eq:trace}
\tau &=& h^{ab} \tau_{ab} \nonumber \\
&=& - \gamma \chi^1 \sqrt{-\mu} \left[2D_c u^c + (d-1) \nabla_c u^c\right] + (d+1)(d-1)\gamma \chi^1 \frac{\dot{\mu}}{2\mu} \nonumber \\
&=& - \gamma \chi^1 \sqrt{-\mu} (d+1) \nabla_c u^c + (d+1)(d-1)\gamma \chi^1 \frac{\dot{\mu}}{2\mu} \nonumber \\
&=& - (d+1) \gamma \chi^1 \left[\sqrt{-\mu} \nabla_c u^c - (d-1) \frac{\dot{\mu}}{2\mu}\right]\;,
\end{eqnarray}
where we have used the fact that $\nabla_a u^a = D_a u^a$, since $u^b\nabla_a u^b = 0$. The traceless part of $\tau_{ab}$ is thus given by
\begin{eqnarray} \label{eq:shear}
\tilde{\tau}_{ab} &=& \tau_{ab} - \frac{\tau}{d-1} h_{ab} \nonumber \\
&=& - 2 \gamma \chi^1 \sqrt{-\mu} \left[D_{(a}u_{b)} - \frac{1}{d-1}\nabla_c u^c\; h_{ab}\right].
\end{eqnarray}
We recognize in the above expression that $\tilde{\tau}_{ab}$ is proportional to the shear 
\begin{eqnarray}\label{shearr}
\sigma_{ab} &:=& D_{\langle a}u_{b \rangle} \nonumber \\
&=& D_{(a}u_{b)} - \frac{1}{d-1}\nabla_c u^c\; h_{ab}
\end{eqnarray}
of a dissipative fluid with velocity $u^c$; namely,
\[
\tilde{\tau}^{ab} = - 2\gamma \chi^1 \sqrt{-\mu}\; \sigma^{ab}.
\]

Now, let us check the resulting
conformal weight orders of the different terms involved. Notice the right hand side of 
equation (\ref{tau}) implies $- \tau_{ab}/\gamma$ has conformal weight $(2+d) - 1 -1  = d$, as desired. 
Recall also that, in a gradient expansion sense, the first order of the stress energy tensor, i.e. in
terms of only first derivatives, is given by 
\begin{equation}
T^{(1)}_{ab} = 2\chi^1\tilde{\tau}_{ab} = -4\gamma (\chi^1)^2 \sqrt{-\mu}\; \sigma_{ab} ,
\end{equation}
while in the standard treatment in the literature (see, for instance, \cite{baier2008relativistic,0264-9381-27-2-025006}) has this order defined as
\[
T^{(1)}_{ab} = - 2 \eta\; \sigma_{ab}.
\]
Thus, defining
\[
\gamma :=  \frac{\eta}{2(\chi^1)^2\sqrt{-\mu}}\;,
\]
we arrive at the same result.

\subsection{Landau frame at first order}

In this section we compute the Landau frame at first order, that is, considering only terms that are up to first order in the dissipative tensor $\xi_{ab}$.  To this order, we have:
\begin{eqnarray}\label{fullT1}
T^{ab} &=& T_o^{ab} + T_1^{ab} \nonumber \\
&=& \left(4\chi^o_{\mu\mu} + \chi^1\frac{\nu}{\mu^2}\frac{d^2(d+1)}{d-1}\right)\left[\xi^a\xi^b - \frac{\mu}{d}g^{ab}\right] - \frac{4(d+1)\chi^1}{\mu}\xi^{(a}r^{b)}\nonumber \\
&& \hspace{8.3cm} + 2\chi^1\tilde{\tau}^{ab} \nonumber \\
&=& T_A{}^{ab} + T_B{}^{ab},
\end{eqnarray}
with
\[
T_A{}^{ab} := \left(4\chi^o_{\mu\mu} + \chi^1\frac{\nu}{\mu^2}\frac{d^2(d+1)}{d-1}\right)\left[\xi^a\xi^b - \frac{\mu}{d}g^{ab}\right]\;,
\]
and
\[
T_B{}^{ab} := - \frac{4(d+1)\chi^1}{\mu}\xi^{(a}r^{b)} + 2\chi^1\tilde{\tau}^{ab}.
\]
To identify the Landau frame, we look for a time--like vector field $U^a$ such that $U^a U_a = -1$ and
\begin{equation}
T^a{}_b U^b = \lambda U^a,
\end{equation}
for some real $\lambda$. We propose the ansatz
\begin{equation}
U^a = \alpha \xi^a + \beta r^a,
\end{equation}
with $\alpha$ and $\beta$ some real functions to be determined. 
Contracting with the two contributions to the stress energy tensor we get,
\[
T_A{}^{ab}U_b = \alpha\left(d(d+1)\chi^1\frac{\nu}{\mu} - 2d\chi^o_{\mu}\right)\xi^a + 2\beta\chi^o_{\mu}r^a + \mbox{2nd. order terms},
\]
and
\[
T_B{}^{ab}U_b = -2\alpha (d+1)\chi^1 r^a + \mbox{2nd. order terms},
\]
and so, up to first order,
\begin{eqnarray}
T^{ab}U_b &=& \alpha\left(d(d+1)\chi^1\frac{\nu}{\mu} - 2(d-1)\chi^o_{\mu}\right)\xi^a + \left(2\beta\chi^o_{\mu} -2\alpha (d+1)\chi^1\right) r^a \nonumber\\
&=& \lambda\alpha\xi^a + \lambda\beta r^a. 
\end{eqnarray}
The above equality yields
\begin{eqnarray}
\lambda &=& d(d+1)\chi^1\frac{\nu}{\mu} - 2(d-1)\chi^o_{\mu} \nonumber \\
&=& - \rho_1 - \rho_o \nonumber \\
&=& - \rho, 
\end{eqnarray}
as expected, since it is the eigenvalue of the unique time--like eigenvector of the full $T^{ab}$ at this order. The remaining equation yields
\begin{equation} \label{alpha-beta-rel}
\left[2d\chi^o_{\mu} - \frac{d(d+1)\chi^1\nu}{\mu}\right]\beta = 2(d+1)\chi^1\alpha.
\end{equation}
Since $U^a$ is time-like and unitary,
\begin{eqnarray}
-1 &=& U^a U_a \nonumber \\ 
&=& \alpha^2\mu + \beta^2 r^a r_a,
\end{eqnarray}
from which we get, to first order, that $\alpha = 1/\sqrt{-\mu}$ (we discard the negative solution for $\alpha$ so that $U^a$ has the same orientation as $\xi^a$). The explicit value of $\beta$ can be obtained from 
the relation (\ref{alpha-beta-rel}). Thus, the heat flux defined as the component of $T^a{}_b U^b$ that is orthogonal to $U^a$ is obviously zero, but nevertheless, the corresponding component of $S^a$ that is orthogonal to $U^a$ is actually non--vanishing, for which there is always a way in which thermal energy is going away.

\section{Second order theory}
\label{sec6}

In this section we tackle the second order contribution, characterizing in a complete way the class of conformal second order dissipative fluid theories. Accounting for this order is important as it will be crucial to
establish symmetric hyperbolicity of the underlying system and thus allowing for well posedness of relevant
problems. The hyperbolicity analysis will be carried out in the next section using the second order results
derived next.

To compute the second order quantities, we make use of a number of useful expressions 
detailed in appendix \ref{appendix2nd}. Recall that strictly at second order we have
\begin{equation} \label{eq:chi2}
\chi^2 = \sum_{i=1}^3{\chi^2{}_i\; \psi_i},
\end{equation}
thus, 
\begin{equation}
T_2^{ab} = \sum_{i=1}^3{T_{2i}^{ab}}, 
\end{equation}
where, for $i=1,2,3$,
\begin{equation}
 T_{2i}^{ab} = 2\psi_i \left(2 \chi^2_{i \mu \mu} \xi^a \xi^b+ \chi^2_{i\mu} g^{ab}  \right)
+ 2 \chi^2_{i\mu} \left(\xi^a \frac{\partial \psi_i}{\partial \xi_b} + \xi^b \frac{\partial \psi_i}{\partial \xi_a} \right)
+ \chi^2_i  \frac{\partial^2 \psi_i}{\partial \xi_a \partial \xi_b}\,,
\end{equation}
with $\chi^2_{i \mu} := \frac{d\chi^2{}_i}{d\mu}$. The corresponding trace is composed of three contributions, namely
\begin{equation}
g_{ab}T_2{}^{ab} = \sum_{i=1}^3{g_{ab}T_{2i}^{ab}},
\end{equation}
with
\begin{equation}
g_{ab}T_{2i}^{ab} = 4\psi_{i}\left(\mu \chi_{i\mu\mu}^{2} + \frac{d}{2} \chi_{i\mu}^{2}\right) + 4 \chi_{i\mu}^{2} \xi^{a}\frac{\partial \psi_{i}}{\partial \xi^a} + \chi_{i}^{2} g^{ab}\frac{\partial^2 \psi_{i}}{\partial \xi^{a} \partial \xi^{b}}
\end{equation}

Since the three scalars $\psi_i$ are independent, the trace-free condition provides now three independent equations, namely
\begin{eqnarray}
\mu \chi_{3\mu\mu}^{2} + D_4 \chi_{3\mu}^{2} &=& 0  \label{st:3}\\
\mu \chi_{2\mu\mu}^{2} + D_2 \chi_{2\mu}^{2} + 2 \chi_{3}^{2} &=& 0   \label{st:2}\\
\mu \chi_{1\mu\mu}^{2} + D_0 \chi_{1\mu}^{2} + \frac{1}{2} \chi_{2}^{2} &=& 0\;, \label{st:1}
\end{eqnarray}
where,
\[
D_j := \frac{d}{2} + j\;, \qquad j=0,1,2,\cdots
\]

To express the constitutive tensor $A_2^{abc}$ we first write it as,
\begin{eqnarray}
A_2^{abc} &=& \frac{\partial^2 \chi^2}{\partial \xi_a \partial \xi_{bc}}\nonumber \\ 
&=& 2\chi^2_{i\mu} \xi^a \frac{\partial \psi_i}{\partial \xi_{bc}}  + \chi^2_i \frac{\partial^2 \psi_i}{\partial \xi_a \partial \xi_{bc}}.
\end{eqnarray}
By using the identities detailed in appendix \ref{appendix2nd} we arrive at,
\begin{eqnarray*} \label{A2form1}
A_2^{abc} &=& 4 \xi^a \left[\chi^2_{1\mu} \xi^{bc} +  \chi^2_{2\mu} \left(\xi^{(b} \ell^{c)} - \frac{\nu}{d} g^{bc}\right) +  \chi^2_{3\mu} \nu\left(\xi^b \xi^c - \frac{\mu}{d} g^{bc}\right)\right] \\
&+& 2\chi^2_2 \left(g^{a(b} \ell^{c)} + \xi^{(b} \xi^{c)a} - \frac{2}{d} g^{bc} \ell^{a}\right) \\
&+& 4 \chi^2_3 \left[\ell^{a} \left(\xi^b \xi^c - \frac{\mu}{d} g^{bc}\right) + \nu\left(g^{a(b} \xi^{c)} - \frac{1}{d}\xi^a g^{bc}\right)\right] \, .
\end{eqnarray*}

Now, let us assess the conditions imposed by conformal invariance. For such analysis, we find it convenient
to make use of equations  (\ref{A1}, \ref{A2}, \ref{A3}) which allow us to write the more general tensor field of rank $(3,0)$ symmetric in the last two indices that depends algebraically on the metric and the variables $(\xi_a, \xi_{ab})$ up to quadratic terms in $\xi_{ab}$.  With this, we then impose that the coefficients satisfy the conformal invariance condition. Specifically, we can express:
\begin{eqnarray}
A_2^{abc} &=& \mathcal{A}\frac{\nu}{\mu}\left(\xi^a g^{bc} + 2 g^{a(b} \xi^{c)} - \frac{(d+2)}{\mu} \xi^a\xi^b\xi^c\right) \nonumber \\
                 & & + \mathcal{B}\left[\xi^a \xi^{bc} + 2 \xi^{a(b} \xi^{c)} - \frac{2}{\mu} \left(2 \xi^a \ell^{(b} \xi^{c)} + \ell^a \xi^b\xi^c\right) + \frac{4\nu}{\mu^2} \xi^a\xi^b \xi^c\right] \nonumber\\
                 & &+  \mathcal{C}\left[\ell^a g^{bc} + 2 g^{a(b} \ell^{c)} - \frac{\nu}{\mu}\left(\xi^a g^{bc} + 2 g^{a(b} \xi^{c)}\right)\right. \nonumber\\
                 & & - \left.\frac{(d+1)}{\mu}\left(\ell^a \xi^b \xi^c + 2 \xi^a \ell^{(b} \xi^{c)}\right) + 3(d+1)\frac{\nu}{\mu^2} \xi^a \xi^b \xi^c\right]\nonumber \\
                 &=& \left(\mathcal{A} - \mathcal{C}\right)\frac{\nu}{\mu}\left(\xi^a g^{bc} + 2 g^{a(b} \xi^{c)}\right) + \mathcal{B}\left(\xi^a \xi^{bc} + 2 \xi^{a(b} \xi^{c)}\right) \nonumber\\
                 & & + \mathcal{C}\left(\ell^a g^{bc} + 2 g^{a(b} \ell^{c)}\right) - \frac{2\mathcal{B} + (d+1)\mathcal{C}}{\mu} \left(\ell^a \xi^b \xi^c + 2 \xi^a \ell^{(b} \xi^{c)}\right)\nonumber\\
                 & & + \left[-(d+2)\mathcal{A} + 4\mathcal{B} + 3(d+1) \mathcal{C}\right]\frac{\nu}{\mu^2} \xi^a\xi^b\xi^c \label{A2form2}
\end{eqnarray}     
Inspection of equation (\ref{A2form2}) implies the following relations,
\begin{eqnarray*}
\mathcal{B} &=& 4\chi^2_{1\mu} = \chi^2_2 \\ 
\mathcal{C} &=& \mathcal{B} = -\frac{4}{d}\left(\chi^2_2 + \mu \chi^2_3\right) \\
\mathcal{A}-\mathcal{C} &=& -\frac{4\mu}{d}\left(\chi^2_{2\mu} + \mu\chi^2_{3\mu} + \chi^2_3\right) = 2\mu\chi^2_3 \\
2 \mathcal{B} + (1+d) \mathcal{B} &=& 4\mu \chi^2_3 = 2 \mu \chi^2_{2\mu} \\
-(d+2)\mathcal{A} + 4\mathcal{B} + 3(d+1) \mathcal{C} &=& 4\chi^2_{3\mu} \mu^2
\end{eqnarray*}
and, in turn,
\begin{eqnarray} \label{generaleqns}
&& \chi^2_2 = 4 \chi^2_{1\mu} \label{chi21} \\
&& \chi^2_3 = \frac{1}{2}\chi^2_{2\mu} \label{chi23} \\
&& \mu\chi^2_{3\mu} + D_3\chi^2_{3} = 0 \, .
\end{eqnarray}
The conditions so derived are compatible with equations (\ref{st:3}), (\ref{st:2}) and (\ref{st:1}), which now become a decoupled system of ordinary second order equations for the unknowns $\chi^2_i$; namely

\begin{eqnarray}
\mu \chi_{3\mu\mu}^{2} + D_4 \chi_{3\mu}^{2} &=& 0  \label{asta:3}\\
\mu \chi_{2\mu\mu}^{2} + D_3 \chi_{2\mu}^{2} &=& 0   \label{asta:2}\\
\mu \chi_{1\mu\mu}^{2} + D_2 \chi_{1\mu}^{2} &=& 0\;,  \label{asta:1}
\end{eqnarray}
which clearly show that one is proportional to the derivative of the 
previous, i.e., $\chi^2_{i+1} \propto \chi^2_{i\mu}$, $i=1,2$,
in agreement with what we already found from equations (\ref{st:3}), (\ref{st:2}) and (\ref{st:1}).
Notice also that $A_2^{abc}$ is trace--free in all entries.
Also, from the above relations we get
\[
\mathcal{B} = \mathcal{C}  = \chi^2_2\;,\qquad \mathcal{A} = -D_1 \mathcal{B}. 
\]

\subsection{Second order contribution to stress-energy tensor}

We begin by noticing that the \textit{unique} solutions to equations (\ref{asta:3}), (\ref{asta:2}) and (\ref{asta:1}), subject to the conformal invariance requirements (\ref{chi21}) and (\ref{chi23}) are
\begin{eqnarray}\label{eq:inv-solns-2ndorder}
\chi^{2}_{1} &=& \chi^2_{1o} + \frac{\chi^2_{o}}{\mu^{D_1}}\;, \nonumber \\
\chi^2_{2} &=& -\frac{4 D_1}{\mu}\left(\chi^{2}_{1} - \chi^2_{1o}\right)\;, \\
\chi^2_{3} &=& \frac{2 D_1 D_2}{\mu^2}\left(\chi^{2}_{1} - \chi^2_{1o}\right)\,; \nonumber 
\end{eqnarray}
with $\chi^2_{1o}$ and $\chi^2_{o}$ real parameters to be determined. Moreover, without loss of generality it is possible to set $\chi^2_{1o}=0$. In fact, introducing the functional $X[\psi_1, \psi_2, \psi_3]$ given by
\begin{eqnarray}
X[\psi_1,\psi_2,\psi_3] &:=& \psi_1 - \frac{4D_1}{\mu}\psi_2 + \frac{2D_1D_2}{\mu^2}\psi_3 \nonumber \\
&=&\xi_{ab}\xi^{ab} - 4\frac{D_1}{\mu} \ell_a \ell^a + \frac{2 D_1 D_2}{\mu^2} \nu^2\;,
\end{eqnarray}
and using the solutions (\ref{eq:inv-solns-2ndorder}), a straightforward calculation shows that the generating function (\ref{eq:chi2}) can be expressed as
\[
\chi^2 = \chi^2_1\;X + \chi^2_{1o}\left[\frac{4D_1}{\mu}\psi_2 - \frac{2D_1D_2}{\mu^2}\psi_3\right].
\]
Thus, using these expressions it is possible to write the second order contribution of the stress-energy tensor in the following form:
\begin{equation}
T^{ab}_2 = \frac{2D_1 X\chi^2_o}{\mu^{D_3}}\left(2D_2 \xi^a \xi^b - \mu g^{ab}\right) - \frac{4 D_1 \chi^2_o}{\mu^{D_2}} X^{(a} \xi^{b)} + \frac{\chi^2_o}{\mu^{D_1}} X^{ab},
\end{equation}
where
\begin{eqnarray}
X^a \equiv \frac{\partial X}{\partial \xi_a} &=& \frac{8 D_1}{\mu^2}\left(\ell^2 \xi^a + D_2 \nu r^a - \mu \xi^{ac}\ell_c \right) \,; \\
X^{ab} \equiv \frac{\partial^2 X}{\partial \xi_a \partial \xi_b} &=& \frac{8 D_1}{\mu^3} \left[- \mu^2 \xi^{ac}\xi^{b}{}_c + 4 \mu \xi^{(a} \xi^{b) c} \ell_c + \ell^2 \mu g^{ab}  - 4 \ell^2 \xi^a \xi^b \right. \nonumber \\
& & \hspace{-1.7cm} + \left. D_2\left(- 8 \nu \xi^{(a} \ell^{b)}- \nu^2 g^{ab} + 6\frac{\nu^2}{\mu} \xi^a\xi^b + 2 \mu \ell^a \ell^b +  \nu \mu \xi^{ab}\right) \right]
\end{eqnarray}
and $\ell^2 := g_{ab}\ell^a\ell^b$. So that nothing depends on the additive constant $\chi^2_{1o}$, and thus we can set it to zero. With these definitions, it is straightforward to verify that
\[
X^a \xi_a =0, \quad X^{ab} \xi_a = - X^b, \quad g_{ab} X^{ab} = -\frac{8D_1}{\mu} X,
\]
which, as expected, imply that
\[
g_{ab}T_2{}^{ab}=0\, .
\]

Let us now inspect what information this second order contribution of the energy--momentum tensor can give us. For  that, let us express it in a more suitable form. First, it is possible to decompose the tensor $X^{ab}$ in the form
\begin{equation} \label{eq:X-deco}
X^{ab} = - \frac{2}{\mu}X^{(a}\xi^{b)} + Y^{ab}\;,
\end{equation}
with $Y^{ab} = Y^{(ab)}$ and $Y^{ab}\xi_a = 0$. In fact, we get
\begin{eqnarray}
Y^{ab} &=& \frac{8D_1}{\mu^3}\left[2\mu\ell_c\xi^{c(a}\xi^{b)} + \left(\frac{4\nu^2}{\mu}D_2 - 2\ell^2\right)\xi^a\xi^b - 6D_2\nu\ell^{(a}\xi^{b)} \right. \nonumber \\ 
&-& \left. \mu^2\xi^{ac}\xi^b{}_c + \left(\mu\ell^2 - D_2\nu\right)g^{ab} + 2D_2\mu\ell^a\ell^b + D_2\mu\nu\xi^{ab}\right].
\end{eqnarray}
With a relation like (\ref{eq:X-deco}), we get
\begin{equation} \label{T2contr}
T_2^{ab} = \frac{\chi^2_1 X}{\mu^2}\frac{d(d+2)(d+3)}{d-1}\left[\xi^a\xi^b - \frac{\mu}{d}g^{ab}\right] - \frac{2(d+3)\chi^2_1}{\mu}X^{(a}\xi^{b)} + \chi^2_1 \tilde{Y}^{ab},
\end{equation}
where, as before, we have denoted by $\tilde{Y}^{ab}$ the traceless part of $Y^{ab}$. In a similar way to the previous orders, we get the second order contributions of density and presure of the fluid, namely,
\begin{equation}
\rho_2 := -(d+2)(d+3)\frac{\chi^2_1 X}{\mu}\;, \qquad p_2 := -\frac{(d+2)(d+3)}{d-1}\frac{\chi^2_1 X}{\mu},
\end{equation}
which clearly satify the conformal equation of state $
\rho_2 = (d-1)p_2$.

\subsection{Constitutive relations up to second order}

We now present the most general constitutive equation that arises from (\ref{dissipeq}) at second order. This equation may be used to derive constitutive relations like the one found in (\ref{fourierlaw}). Nevertheless, and for future usage, we shall only write down the corresponding contribution at this order, as well as for a complete presentation of the theory considering second order contributions.

Using the conformally invariant solutions (\ref{eq:inv-solns-2ndorder}), the second order contribution to the constitutive tensor $A_2^{abc}$ in equation (\ref{A2form1}) reads
\begin{eqnarray} \label{eq:A2-finalform}
A_2^{abc} &=& \frac{4D_1\chi^2_1}{\mu^3} \xi^a \left[-\mu^2 \xi^{bc} + 4D_2\mu \left(\xi^{(b} \ell^{c)} - \frac{\nu}{d} g^{bc}\right) -2D_2D_3 \nu \left(\xi^b \xi^c - \frac{\mu}{d} g^{bc}\right)\right] \nonumber \\
&-& \frac{8D_1\chi^2_1}{\mu}\left[g^{a(b} \ell^{c)} + \xi^{(b} \xi^{c)a} - \frac{2}{d} g^{bc} \ell^{a}\right] \nonumber \\
&+& \frac{8D_1D_2\chi^2_1}{\mu^2} \left[\ell^{a} \left(\xi^b \xi^c - \frac{\mu}{d} g^{bc}\right) + \nu\left(g^{a(b} \xi^{c)} - \frac{1}{d}\xi^a g^{bc}\right)\right]\, .
\end{eqnarray}

After a tedious but rather straightforward calculation, the covariant derivative of $A^{abc}$ is
\begin{eqnarray} \label{diva2ndorder}
\nabla_a A_2^{abc} &=& \frac{8D_1D_2D_3\chi^2_1\mathcal{G}}{\mu^4}\left(\xi^b \xi^c - \frac{\mu}{d}g^{bc}\right) + \frac{4D_1\chi^2_1}{\mu^2}(D_2\dot{\mu} - \mu D)\;\xi^{bc}\nonumber\\
&+& \frac{16D_1D_2\chi^2_1}{\mu^3}(\mu D - \dot{\mu}D_3)\left(\xi^{(b}\ell^{c)} - \frac{\nu}{d}g^{bc}\right) - \frac{4D_1\chi^2_1}{\mu}\dot{\xi}^{bc}\nonumber\\
&+& \frac{16D_1D_2\chi^2_1}{\mu^2}\left(\dot{\xi}^{(b}\ell^{c)} + \xi^{(b}\dot{\ell}^{c)} - \frac{\dot{\nu}}{d}g^{bc}\right) - \frac{8D_1D_2D_3\chi^2_1}{\mu^3}\nu\left(2\dot{\xi}^{(b}\xi^{c)} - \frac{\dot{\mu}}{d}g^{bc}\right)\nonumber\\
&+& \frac{8D_1D_2\chi^2_1}{\mu^2}\left(\ell^{(b}\nabla^{c)}\mu + \xi^{(b}\xi^{c)a}\nabla_a\mu - \frac{2\ell^a\nabla_a\mu}{d} g^{bc}\right) \nonumber\\
&-& \frac{8D_1\chi^2_1}{\mu}\left[\nabla^{(b}\ell^{c)} + \xi^{a(b}\nabla_a \xi^{c)} + \xi^{(b}\nabla_a \xi^{c)a} - \frac{2\nabla_a \ell^a}{d}g^{bc}\right] \nonumber \\
&-& \frac{8D_1D_2D_3\chi^2_1}{\mu^3}\nu\left[\xi^{(b}\nabla^{c)}\mu - \frac{\dot{\mu}}{d}g^{bc}\right] + \frac{8D_1D_2\chi^2_1}{\mu^2} \mathcal{G}^{bc}\;,
\end{eqnarray}
where $\dot{\xi}^{ab}:=\xi^c\nabla_c\xi^{ab}$, $\dot{\xi}^c:=\xi^a\nabla_a \xi^c$, and $\mathcal{G}$ and $\mathcal{G}^{bc}$ are given by
\[
\mathcal{G}:=\nu\left(D_4\dot{\mu} - D\mu\right) - \mu \left(\dot{\nu} - \frac{\mu}{D_3}\nabla_a\ell^a + \ell^a\nabla_a\mu\right);
\]
\[
\mathcal{G}^{bc} := 2\ell^a \xi^{(b}\nabla_a \xi^{c)} + \xi^{(b}\nabla^{c)}\nu + \nu\nabla^{(b}\xi^{c)} - \frac{1}{d} \left(\ell^a\nabla_a\mu + \dot{\nu} + D\nu\right) g^{bc}.
\]

In complete analogy with the way we proceeded to first order, contraction with $\xi_c$ and projection onto the perpendicular space in the remainding index leads to the identity
\begin{eqnarray} \label{eq:perp-NablaA_2}
h^d{}_b\xi_c\nabla_a A_2^{abc} &=& - \frac{2(d^2 + 4d + 8)D_1\chi^2_1\nu}{\mu^2} h^d{}_b\dot{\xi}^b + \frac{4(d+1)D_1\chi^2_1}{\mu} h^d{}_b \dot{r}^b \nonumber \\
&+& \frac{4D_1\chi^2_1}{\mu}\left(\frac{D_3\dot{\mu}}{\mu} - D\right) r^d + \frac{4D_1\chi^2_1}{\mu} \tau^{dc}\dot{\xi}_c - \frac{4D_1D_2^2\chi^2_1\nu}{\mu^2} D^d\nu \nonumber \\
&+& \frac{4D_1^2\chi^2_1}{\mu}\tau^{dc}D_c\mu + \frac{8D_1^2\chi^2_1}{\mu} h^d{}_b r^a \nabla_a \xi^b + \frac{4D_1D_2\chi^2_1}{\mu} D^d\nu \nonumber \\
&-& \frac{4D_1\chi^2_1}{\mu} h^d{}_b\xi_c\nabla^b\ell^c - 4D_1\chi^2_1 h^d{}_b\nabla_a \tau^{ab}.
\end{eqnarray}

In this case, the constitutive relations should follow by proposing the most general source tensor $I^{bc}$ up to second order in dissipative variables, which may be constructed only as an algebraic function of the fields $\xi^a$, $r^a$, $\tau_{ab}$ and $g_{ab}$. In fact, referring once again to the discussion presented in Appendix \ref{app1}, it can be easily shown that such a tensor has the following form:
\begin{equation}\label{full-source}
I^{bc} = I_o \xi^b \xi^c + I_1 \xi^{(b} r^{c)} + I_2 \xi^{(b}\tau^{c)a}r_a + I_3 h^{bc} + I_4 r^b r^c + I_5 \tau^{bc} + I_
6 \tau^{bd}\tau^c{}_d.
\end{equation}

It is important to remark that (\ref{full-source}) is already the full and more general source tensor for the (full) constitutive equation, that reads
\begin{equation}\label{full-constitutive-relation}
\nabla_a \left(A_1^{abc} + A_2^{abc}\right) = I^{bc},
\end{equation}
without further assumptions of form. The functions $I_j$ ($j = 0, 1, \cdots, 6$) should be fixed in such a way that $I^{bc}$ is trace--free and the entropy production $\sigma = -\xi^{ab}I_{ab}$ satisfies the inequality $\sigma \geq 0$. Also, and in complete analogy with the first order theory, there is an additional requirement for the coefficients, which is related to the global conformal weight of the full constitutive equation (\ref{dissipeq}). Since each one of the contribution terms appearing in (\ref{full-source}) must have the \textit{same} conformal weight, as well as it must be equal to the conformal weight of the left hand side terms of (\ref{dissipeq}), the full constitutive equation (\ref{full-constitutive-relation}) with $I^{bc}$ given by (\ref{full-source}) is sensible in so far that the functions $I_j$ have the appropriate conformal weights. This implies that in order to obtain plausible constitutive relations, one should make sure that
each one of the terms has the same conformal weight, and thus, each $I_j$ might have different conformal weights a priori so that, upon multiplication with the corresponding factors, give the right (and same) conformal weight. 

The issue of finding which transport coefficients emerge from this theory is not direct a priori, since it requires a detailed analysis of each one of the terms that contributes to the divergence of $A^{abc}$. Unlike the first order theory --in which it was possible to obtain the shear transport coefficient just by comparing the shear tensor with the transverse traceless part of the energy-momentum tensor--, a quite different strategy at second order seems to be needed. The coefficients obtained may be compared, for instance, with the ones already derived from the gravity side, via the fluid/gravity correspondence (for references, see \cite{Bhattacharyya:2008mz,baier2008relativistic}). Obtaining specific expressions for these contributions is however outside the scope of the present work.

\subsection{Entropy density contribution}

In analogy with the zeroth and first order contributions of the theory, it is possible to compute the second order correction of the total entropy density. We have

\begin{equation}\label{entropy-2nd}
S_2^a = \frac{\partial \chi^2}{\partial \xi_a} - \xi_b T_2^{ab} - \xi_{bc} A_2^{abc}.
\end{equation}
Let us compute each of the terms that contribute to $S_2^a$. At this point, it is convenient to introduce the functional
\[
G[\psi_1,\psi_2,\psi_3] := \psi_1 - \frac{4D_2}{\mu}\psi_2 + \frac{2D_2D_3}{\mu^2}\psi_3,
\]
and the vector field
\[
G^a := G\xi^a - \frac{2(d+4)\nu}{\mu}\ell^a + 4 \xi^{ab}\ell_b\;.
\]
Using relations (\ref{eq:chi2}), (\ref{eq:inv-solns-2ndorder}), (\ref{T2contr}) and (\ref{eq:A2-finalform}), we get simple expressions for the desired contributions, namely
\[
\frac{\partial \chi^2}{\partial \xi_a} = -\frac{(d+2)\chi^2_1}{\mu}G^a, \qquad \xi_b T_2^{ab} = \frac{(d+2)(d+3)\chi^2_1}{\mu} G^a, 
\]
and
\[
\xi_{bc} A_2^{abc} = - \frac{2(d+2)\chi^2_1}{\mu} G^a.
\]

Finally, the second order contribution to the total entropy density is given by
\begin{equation}\label{entropy2ndFINAL}
S^a_2 = -\frac{(d+2)^2\chi^2_1}{\mu}\;G^a.
\end{equation}

It is possible to compare the above expression with the results found by Hiscock and Lindblom \cite{hiscock1983stability}. In fact, recalling the orthonormal trace--free decomposition for $\xi^{ab}$ given in (\ref{orto-traceless-xi}), and relations (\ref{eq:inv-solns-2ndorder}) one gets
\begin{eqnarray}\label{entropy2ndexplic}
S_2^a &=& -\frac{(d+2)^2\chi^2_o}{\mu^{D_2}}\left(\frac{d^2(d+1)}{2(d-1)}\frac{\nu^2}{\mu^2} - 2(d+1)r^c r_c + \tilde{\tau}^{bc}\tilde{\tau}_{bc}\right)\xi^a \nonumber \\
&+& \frac{4(d+2)^2\chi^2_o}{\mu^{D_3}}\frac{d(d+1)}{2(d-1)}\nu r^a - \frac{4(d+2)^2\chi^2_o}{\mu^{D_2}}\tilde{\tau}^{ab}r_b,
\end{eqnarray}
which can be written as
\begin{equation}\label{entropy-hiscock}
S_2^a = -(\beta_o \Pi^2 + \beta_1 q^c q_c + \beta_2 \pi^{bc}\pi_{bc})\frac{u^a}{2T} + \alpha_o\frac{\Pi q^a}{T} + \alpha_1 \frac{\pi^{ab}q_b}{T},
\end{equation}
with the straightforward identifications
\[
u^a = \frac{\xi^a}{\sqrt{-\mu}}, \quad T = \frac{1}{\sqrt{-\mu}}, \quad \Pi \equiv p_1, \quad \pi^{ab} \equiv 2\chi^1 \tilde{\tau}^{ab},
\]
in which relations and definitions (\ref{rho-p-firstorder}), (\ref{chi1soln}) and (\ref{heatfluxfinal}) were used for $p_1$, $\chi^1$ and $q^a$ respectively. Moreover, the coefficients $\beta_o$, $\beta_1$, $\beta_2$ and $\alpha_o$, $\alpha_1$ emerge by comparing (\ref{entropy2ndexplic}) and (\ref{entropy-hiscock}), yielding
\begin{equation}
\beta_o = \frac{(-1)^{d/2}(d+2)^2(d-1)\chi^2_o}{(\chi^1_o)^2}\frac{1}{T^d},
\end{equation}
and
\begin{equation}
\beta_1 = \frac{\mu}{d^2-1}\beta_o , \quad \beta_2 = \frac{\beta_o}{2(d-1)}, \quad \alpha_o = \alpha_1 = \frac{\beta_o}{1-d^2}.
\end{equation}
These coefficients essentially model all dissipative contributions to the entropy at this order (scalar, vector and tensor ones), as well as couplings between viscosity effects and heat fluxes \cite{rezzolla2013relativistic}. We find it important to stress out that, as a consequence of the conformal invariance requirement, they are not independent.

\section{Well posedness of the full second order theory}
\label{sec7}

The issue of stability and causality in relativistic fluid theories is a crucial aspect that has been the subject of intense scrutiny in the past (e.g. \cite{geroch1990dissipative,geroch1991causal,PhysRevD.81.114039,0954-3899-35-11-115102,calzetta2010linking}). It is well known that the Landau--Eckart theories have an ill--posed initial value formulation, as well as that thermal fluctuations propagate in an non-causal way (more specifically, the linearized modes grow linearly with the frequency). Generally, the concepts of stability, causality and hyperbolicity are independent (for instance, Navier--Stokes equations are stable, but non-causal), and there may be weakly hyperbolic systems that are causal, but not stable.

Here we study the hyperbolicity of the full theory developed along this work; i.e., considering second order contributions of dissipative effects\footnote{Second-order theory is highly successful when confronting experimental results. Considering higher orders in dissipation would induce a greater number of parameters for which there are no experimental estimates. Thus, our theory would be applied only when the neglected higher order terms are small, so that the fluid is near equilibrium.}. We start by briefly reviewing symmetric hyperbolic systems, introducing definitions and some key results we shall make use of. We follow the works \cite{friedrichs1954symmetric,friedrichs1971systems,geroch1996partial,geroch1995relativistic,geroch1991causal}. After that, we show it is possible to choose the arbitrary constant $\chi^2_{1o}$ in the generating function at second order, $\chi^2$, to ensure the system results hyperbolic, and we prove that the full theory up to second order is symmetric hyperbolic near equilibrium states. This implies that the corresponding initial value problem is locally well posed.

\subsection{Hyperbolicity}

As discussed in the previous sections, dissipative relativistic fluids are completely determined by giving a single scalar function $\chi$ that depends smoothly on the variables $(\xi_a,\xi_{ab})$, and a dissipation-source tensor field $I^{ab}$ which is an algebraic function of those variables and the spacetime metric, and stores the net entropy production of the system. In order to see more clearly the structure of the dynamical equations, it is useful to introduce a collective abstract variable in the following way. Let $V$ be the linear space of all fields of the form
\begin{equation}
X^{A} = (V^{a},S^{ab}),
\end{equation}
where $V^a$ and $S^{ab}$ are, respectively, a vector field and a symmetric and traceless tensor field, both defined on $\mathcal{M}$. Here, capital indices stand for the whole set of tensor indices within this collection. Within this abstract vector space we shall study the hyperbolicity of the theory. Let us start with the following
\begin{definition}
A vector $\xi^C\in V$ is called a \textit{fluid state} if $\xi^C:=(\xi^a, \xi^{ab})$, where $\xi^a$ and $\xi^{ab}$ are  solutions to the conservation equations (\ref{Tabcons}) and (\ref{dissipeq}). We shall denote by $V_F$ the set of all fluids states.
\end{definition} 

This is a concept that has already appeared in section \ref{perfect-fluids}. From this, it is possible to re--express the set of dynamical equations (\ref{Tabcons}) and (\ref{dissipeq}) as the first order system given by
\begin{equation} \label{geroch-form-system}
K^a{}_{AB} \nabla_a \xi^{B} = J_{A},
\end{equation}
where
\begin{equation}
K^a{}_{AB} := \frac{\partial^3 \chi}{\partial \xi_a \partial \xi^A \partial \xi^B},
\end{equation}
and $J_A := (0,\;I_{ab})$. At this point, let us observe that, by the symmetries of $\xi^a$ and $\xi^{ab}$, system (\ref{geroch-form-system}) has $d + \frac{d(d+1)}{2}-1$ unknowns, which already coincides with the number of equations. Thus, both spaces have the same dimension. On the other hand, recall that, by construction, the principal part $K^a{}_{AB}$ is \textit{symmetric} in the capital indices; i.e., $K^a{}_{AB} = K^a{}_{(AB)}$, since partial derivatives commute.

\begin{definition}
System (\ref{geroch-form-system}) is \textbf{symmetric hyperbolic} at $\xi^C$ if there exists a covector $t_a$ such that $t_aK^a_{AB}(\xi^C)$ is \textbf{positive definite}.
\end{definition}

Recall that a general quadratic form $h_{AB}(t)$ is \textit{positive definite} if the contraction $h_{AB}(t) X^A X^B$ is positive for any $X^A \in V$ (that is, not only for fluid states). The hyperbolicity condition guarantees that (\ref{geroch-form-system}) has a well posed initial value formulation \cite{friedrichs1954symmetric}.

\subsection{Symmetric hyperbolicity of the conformal fluid theories}

In this section we prove a fundamental theorem about well posedness for these type of conformal fluids, which is one of the main results of this work. Specifically, we show that there exists a symmetrizer for the full theory, and it turns out that it is positive definite in an open subset of fluid states around an equilibrium one. Following Friedrichs--Lax argument regarding hyperbolicity of first order systems \cite{friedrichs1971systems,friedrichs1954symmetric}, we conclude that there exists a well posed initial value problem for these theories around equilibrium states. We now state the main theorem:

\begin{theorem}
\textbf{Symmetric hyperbolicity of the full theory.} Let $(\mathcal{M}, g_{ab})$ be a time-oriented spacetime of arbitrary dimension $d$, and $\xi_a$ be an equilibrium state of the theory. Then, there always exists an open set $\mathcal{O}$ around $\xi_a$ such that system (\ref{geroch-form-system}) is hyperbolic within the whole $\mathcal{O}$.
\end{theorem}
\textit{Proof.} Following Definition 7.2, it is enough to consider $t_a$ to be in the direction of $\xi_a$ and evaluate the quadratic form
\[
h_{AB}(\xi) := \xi_a K^a_{AB}
\]
at \textit{equilibrium states}, that is, on perfect fluid solutions. The key idea is the following: if one manages to show that $h_{AB}(\xi)$ is positive definite, there will be an open set of fluid states $\mathcal{O}$ around it, such that $h_{AB}(t) > 0$ for all $t_a\in\mathcal{O}$. Thus, we shall look for the conditions for $h_{AB}(\xi)$ to be positive definite.

In fact, evaluating $K^a_{AB}$ at equilibrium states, we straightforwardly get
\begin{eqnarray}
\stackbin{o}{K^a}_{AB} = \left( \begin{array}{cc}
	\frac{\partial^3 \chi^o}{\partial \xi_a \partial \xi^b \partial \xi^c} & \frac{\partial^3 \chi^1}{\partial \xi_a \partial \xi^b \partial \xi^{cd}}\\
	\frac{\partial^3 \chi^1}{\partial \xi_a \partial \xi^b \partial \xi^{cd}} & \frac{\partial^3 \chi^2}{\partial \xi_a \partial \xi^{bc} \partial \xi^{de}}
	\end{array}
\right),
\end{eqnarray}
where the symbol ${}^{o}$ over $K^a{}_{AB}$ denotes evaluation at equilibrium states, after taking the corresponding derivatives. Now, let us consider the quadratic form
\[
\stackbin{o}{h}_{AB}(\xi) := \xi_a \stackbin{o}{K^a}_{AB}.
\] 
We first assert that the upper diagonal block of $\stackbin{o}{h}_{AB}$ is positive, since it is the expression of the corresponding hyperbolizer of the perfect fluid theory, which is a symmetric--hyperbolic theory, stable and causal. This is so because the corresponding equation of state is that of a pure \textit{radiation} fluid, which already satisfies the requirements for the system to be symmetric hyperbolic (see, for instance, \cite{geroch1996partial}). Thus, if we find that the lower diagonal block is positive as well, then we can make the whole matrix to be positive definite, just by choosing $\chi^2_{1o}$ large enough and with the appropriate sign, where $\chi^2_{1o}$ is the free parameter in the solutions (\ref{eq:inv-solns-2ndorder}).  It is important to stress here the inclusion of second order dissipative effects; otherwise the lower
diagonal block would be zero and it would be rather impossible to satisfy the sought-after condition.
Let us study, therefore, the lower diagonal block\footnote{Interestingly, a first order formulation which is well posed in a weaker sense
has been presented in \cite{Bemfica:2017wps}, and
contemplates an entropy current that is not unique (as it depends on the initial data for the 
higher order dissipative contributions).}.
 
We find it useful to introduce the second order weighted scalars
\begin{equation}
\tilde{\psi}_i := \frac{\psi_i}{\mu^{i-1}},
\end{equation}
where $\psi_i$ are the three main scalars introduced in the second order theory. These scalars do not depend on the norm of $\xi^a$, namely $\mu$, from which they satisfy
\begin{equation}\label{der-tilde}
\xi_a\frac{\partial\tilde{\psi}_i}{\partial\xi_a} = 0.
\end{equation}
With those scalars, the second order generating function (\ref{eq:chi2}) can be re-expressed as 
\begin{equation}
\chi^{2} = \chi^{2}_{1}\tilde{\psi}_{1} + \mu \chi^2_2 \tilde{\psi}_2 + \mu^2 \chi^2_3 \tilde{\psi}_3.
\end{equation}
Thus, 
\[
\xi_a \frac{\partial \chi^2}{\partial \xi_a} = 2\mu \left[\chi^{2}_{1\mu}\tilde{\psi}_{1} + \left(\mu \chi^2_2\right)_{\mu} \tilde{\psi}_2 + \left(\mu^2 \chi^2_3\right)_{\mu} \tilde{\psi}_3\right],
\]
recalling property (\ref{der-tilde}). Now, using the relation between the second order contributions $\chi^{2}_j$ found in (\ref{chi21}), (\ref{asta:3}), (\ref{asta:1}) and (\ref{eq:inv-solns-2ndorder}), we can express everything in terms of $\chi^2_{1}$ and get 
\begin{equation}\label{lowerblockpos}
\xi_a \frac{\partial \chi^2}{\partial \xi_a} = -\frac{2D_1\chi^2_{1o}}{\mu^{D_1}}\left[\tilde{\psi}_{1} - 4\left(\frac{d}{2}+1\right) \tilde{\psi}_2 + 2\left(\frac{d}{2} + 1\right)\left(\frac{d}{2}+2\right) \tilde{\psi}_3\right].
\end{equation}

Notice that the lower diagonal block we need it to be positive definite is indeed the Hessian matrix of
\[
\xi_a \frac{\partial \chi^2}{\partial \xi_a}.
\]
Thus, symmetric hyperbolicity would follow if we could show this function has a definite sign, so that it Hessian has definite sign as well. This convexity condition already appeared in a very nice work by Lax and Friedrichs \cite{friedrichs1971systems}, in which they show that general conservation laws which admit a convex extension are symmetric hyperbolic. Here we shall prove that this is already the case, by showing that the functional
\[
\mathcal{F}[\tilde{\psi}_1, \tilde{\psi}_2, \tilde{\psi}_3] := \tilde{\psi}_{1} - 4\left(\frac{d}{2}+1\right) \tilde{\psi}_2 + 2\left(\frac{d}{2} + 1\right)\left(\frac{d}{2}+2\right) \tilde{\psi}_3
\]
is positive definite for all values of $\xi_{ab}$. As mentioned, choosing a large enough value for the constant $\chi^2_{1o}$ (with the appropriate sign) one ensures the desired positivity condition is satisfied. In order to see that $\mathcal{F}[\tilde{\psi}_1, \tilde{\psi}_2, \tilde{\psi}_3]$ is positive definite, we consider the decomposition of $\xi_{ab}$ given in (\ref{orthonormal-deco-xiab}), and express any linear combination of the scalars $\tilde{\psi}_i$ as a real linear combination of quadratic (and therefore positive) quantities, namely that the equality
\begin{equation}\label{lineargralcomb}
\alpha \tilde{\psi}_1 +\beta \tilde{\psi}_2 + \gamma \tilde{\psi}_3 =  (\alpha + \beta + \gamma) \frac{\nu^2}{\mu^2} + \frac{2\alpha + \beta}{\mu} r_a r^a + \alpha \tau_{ab} \tau^{ab},
\end{equation}
holds for all $\alpha, \beta, \gamma$. In particular, by choosing the linear combination that defines the functional $\mathcal{F}$; i.e.,
\[
\alpha = 1, \quad \beta = - 2(d+2), \quad \gamma = 2\left(\frac{d}{2} + 1\right)\left(\frac{d}{2}+2\right),
\]
equation (\ref{lineargralcomb}) gets
\[
\tilde{\psi}_{1} - 4\left(\frac{d}{2}+1\right) \tilde{\psi}_2 + 2\left(\frac{d}{2} + 1\right)\left(\frac{d}{2}+2\right) \tilde{\psi}_3 =
\]
\begin{equation}
\left(1+d+\frac{d^2}{2}\right)\frac{\nu^2}{\mu^2} -\frac{2(d+1)}{\mu}r_a r^a + \tau_{ab}\tau^{ab}, 
\end{equation}
which is always positive. This concludes the proof.
\begin{flushright}
$\Box$
\end{flushright}

\section{Concluding remarks}
\label{sec8}

In this work we fully characterized the class of conformal dissipative relativistic fluids in arbitrary dimensions, considering dissipative effects up to second order. The dynamical variable is the energy--momentum tensor associated with the fluid, and no conservation of particle numer density is required. We followed the formalism of divergence-type theories due to R. Geroch, L. Lindblom and Pennisi \cite{geroch1990dissipative,pennisi87} from which it is possible to characterize a fluid theory solely by prescribing a generating scalar function and the corresponding dissipation source. Of course, one may wonder what happens if the divergence-type nature of the dynamical equations is relaxed, and instead of that, allow for more general equations; namely, equations which do not represent conservations laws. Certainly this can be done, and perhaps there would be many more available theories with intriguing properties. Nevertheless, one drawback of such enterprise is the difficulty to describe shocks effects, i.e., solutions which starting with smooth initial data produce discontinuities at finite evolution time. Unfortunately we can only make sense of this solutions when their equations represent conservations laws, that is divergence-type theories \cite{geroch1991causal}.

In the present case of conformal dissipative fluids up to second order, we found a \textit{unique} three-parameter family of scalar functions that generate the whole class. We studied the equilibrium states of these theories, and
showed that they are in one-to-one correspondence with the conformal Killing
vector fields of the background metric.

At first order in dissipation, we recovered the standard expressions.
Without assuming any preferred frame (e.g. the Landau or Eckart ones),
we found both heat flux (interpreted it as the
failure of a fluid solution to be in equilibrium) and shear tensor to be non-vanishing. Moreover, even if they start vanishing at the initial time, they are in general generated during evolution. We derived a Fourier-like
equation for the heat flux, as a consequence of imposing the constitutive equations at
the corresponding order. Additionally, We also found the expected relation between the
part of the energy momentum tensor that is completely perpendicular to
the four-velocity of the fluid and the corresponding shear.

Finally, we studied the initial value problem of these theories and the
issue of hyperbolicity. Given that the second order contribution to the the-
ory is fixed up to a constant parameter, it is by no means trivial that the
theory would end up being hyperbolic. We found however that by chosing
it large enough, the theory does turn out to be symmetric-hyperbolic near
equilibrium solutions, which directly implies that it has a locally well-posed
initial value problem.

Looking ahead, there certainly remain some aspects to develop further. 
One of them is related to the
understanding on how the different transport coefficients at second order
emerge from the corresponding constitutive equation in these theories, without assuming any preferred frame. One other point, which is most intriguing,
arises through the fluid/gravity correspondence. Since there is a one-
to-one correspondence between conformal fluids and particular geometries
--which are solutions to Einstein’s field equations-- and that we have seen
conformal fluids can be characterized by a single generating scalar function, 
to what extent can geometries admit a scalar characterization?

\section*{Acknowledgements}
\addcontentsline{toc}{section}{Acknowledgements}

This research work was partially supported by grants PIP and PICT of SeCyT, UNC (O.A.R and M.E.R); as well
as NSERC and CIFAR (L.L).
M.E.R thanks Perimeter Institute for hospitality during part of the development of this work, as well as Stephen Green and Osvaldo Moreschi for discussions. Research at Perimeter Institute is supported by the Government of Canada through the Department of Innovation, Science and Economic Development, Canada, and by the Province of Ontario through the Ministry of Research and Innovation. M.E.R has a doctoral fellowship of CONICET, Argentina.

\appendix
\section{On symmetric traceless tensor decompositions}
\label{app1}

In this appendix we discuss some properties and decomposition of second and third rank tensor fields with the symmetries of $T^{ab}$ and $A^{abc}$ used to describe fluid theories. In particular, we show the orthonormal decomposition of second rank symmetric traceless tensor fields we use throughout this work. 

\subsection{Second rank tensor fields}

We start with second rank tensor fields. Let $S^{ab}$ be a symmetric traceless tensor field, and $V^a$ a vector field with $V^cV_c \neq 0$. Then, there always exist a scalar function $\alpha$, a vector field $R^a$ and a symmetric tensor field $P^{ab}$ satisfying
\begin{equation}\label{symmetries}
R^a V_a = 0, \qquad P^{ab}V_b = 0, \qquad P^{ab}g_{ab} = 0
\end{equation}
such that
\begin{equation}\label{orthonormal-deco}
S^{ab} = \alpha\left(V^a V^b - \frac{V^c V_c}{d}g^{ab}\right) + \frac{2}{V^c V_c} V^{(a}R^{b)} + P^{ab}. 
\end{equation}
In fact, contracting (\ref{orthonormal-deco}) with $V_a V_b$ we get
\[
S^{ab}V_aV_b = \frac{d-1}{d}\;\alpha\;(V^cV_c)^2,
\]
from which
\begin{equation}\label{alpha-deco}
\alpha = \frac{d}{d-1}\frac{S^{ab} V_a V_b}{(V^c V_c)^2}.
\end{equation}
Now, contracting (\ref{orthonormal-deco}) with $V_b$ we obtain
\[
S^{ab}V_b = \frac{d-1}{d}\;\alpha\;V^c V_c\;V^a + R^a,
\]
from which we directly get
\[
R^a = S^{ab}V_b - \frac{S^{cd} V_c V_d}{V^c V_c} V^a.
\]
Notice that $R^a V_a = 0$ as assumed. Finally, the desired tensor $P^{ab}$ is given by
\[
P^{ab} = S^{ab} - \frac{d}{d-1}\frac{S^{cd} V_c V_d}{(V^c V_c)^2} \left(V^a V^b - \frac{V^c V_c}{d}g^{ab}\right) - \frac{2}{V^c V_c} V^{(a}R^{b)}.
\]
Indeed, it is straightforward that $P^{ab}V_b = 0$ and $P^{ab}g_{ab} = 0$ since $S^{ab}$ is traceless by hypothesis. Decomposition (\ref{orthonormal-deco}) is used along this work, by choosing $S_{ab} = \xi_{ab}$ and $V_a = \xi_a$, namely
\begin{equation}\label{orto-decoxi}
\xi_{ab} = \frac{\nu}{\mu^2}\xi_a \xi_b + \frac{2}{\mu}\xi_{(a}r_{b)} + \tau_{ab}.
\end{equation}
Introducing the traceless part of $\tau_{ab}$ as
\begin{equation}\label{traceless-tau}
\tilde{\tau}_{ab} := \tau_{ab} + \frac{\nu}{\mu (d-1)} h_{ab},
\end{equation}
expression (\ref{orto-decoxi}) becomes
\begin{equation}\label{orto-traceless-xi}
\xi_{ab} = \frac{d}{d-1}\frac{\nu}{\mu^2}\left(\xi_a\xi_b - \frac{\mu}{d}g_{ab}\right) + \frac{2}{\mu}\xi_{(a}r_{b)} + \tilde{\tau}_{ab}.
\end{equation}

\subsection{Third rank tensor fields}

Let $S^{ab}$ be any symmetric tensor field, and $V^a$ a vector field. After some inspection, it is possible to assert that the most general tensor $A^{abc}$ with the requiring properties for it to be a constitutive tensor field for a fluid theory (i.e., satisfying condition (\ref{Acondition})) must be of the form
\begin{equation}
A^{abc}= V^{a}S^{bc} + 2S^{a(b} V^{c)}, \label{Atensor}
\end{equation}
if and only if $S^{ab}$ and $V^a$ satisfy the condition
\begin{equation}
2 S^{ab}V_a + S^{ac}g_{ac} V^b = 0.
\end{equation}

In the context of this work, since we make use of some relevant variables $(\xi_a, \xi_{ab})$, tensors $V^a$ and $S^{ab}$ are
to be expressed as algebraic functions of them, at each order. Let us inspect the possibilities that arise at each order. Since at non-dissipative orders $\xi^{ab} = 0$, we can only construct such tensors in terms of $\xi^a$ and $g^{ab}$, namely,
\begin{equation}
 V^a = \xi^a, \qquad S^{ab} = S_o\left(g^{ab} - \frac{d+2}{3 \mu}\xi^a\xi^b\right)\;,
\end{equation}
where $S_o$ is a constant that should be fixed depending on the dissipation order. Indeed, for perfect fluids one should take $S_o = 0$, while for first order theories, $S_o = 1$ as pointed out in equation (\ref{Sab-1storder}) of section \ref{1storder-section}.

For the next order piece, there are three possible terms which are at most linear en the dissipative 
variables $\xi^{ab}$. In fact, and recalling the dissipative vector $\ell^a:= \xi^{ab} \xi_b$ introduced before, those terms are found to be
\begin{eqnarray}
V^a := \xi^a, \qquad  S^{ab} := S_o^1\frac{\nu}{\mu}\left(g^{ab} - \frac{d+2}{3 \mu}\xi^a\xi^b\right)\;, \label{A1} \\ 
V^a := \xi^a, \qquad S^{ab} := S_o^2\left(\xi^{ab} - \frac{2}{\mu}r^{(a} \xi^{b)} + \frac{\nu}{\mu^2} \xi^a\xi^b + \frac{\nu}{\mu(d-1)}h^{ab}\right), \label{A2}\\
V^a := r^a, \qquad S^{ab} := S_o^3\left(g^{ab} - \frac{d+1}{\mu} \xi^a \xi^b\right), \label{A3}
\end{eqnarray}
where $r^a = \ell_a - \frac{\nu}{\mu}\xi_a$, $h^a{}_b = \delta^a{}_b - \frac{1}{\mu}\xi^a \xi_b$ is the projector onto the orthogonal surfaces of $\xi^a$, as before, and ($S_o^1$, $S_o^2$, $S_o^3$) real constants to determine. Finally, we found it useful to rename the traceless $\xi^a$-completely orthogonal part of $\xi^{ab}$ as

\begin{equation}\label{orthonormal-deco-xiab}
\tilde{\tau}^{ab} := \xi^{ab} - \frac{2}{\mu}r^{(a} \xi^{b)} + \frac{\nu}{\mu^2} \xi^a\xi^b + \frac{\nu}{\mu(d-1)}h^{ab},
\end{equation}
which already coincides with (\ref{traceless-tau}), as expected.

\section{Useful expressions for...}
\label{appendix2nd}

In this appendix we state the main identities needed to do the calculations for the different tensor fields to second order in dissipation. We recall that, when taking derivatives to symmetric and traceless tensor fields, we substract the trace of the corresponding result, obtaining by this way just the symmetric and traceless part of the corresponding derivative field.

\subsection{... second order $T^{ab}$}

\[
\frac{\partial \xi^{ab}}{\partial \xi_{cd}}= g^{a(c}g^{d)b} - \frac{g^{ab}g^{cd}}{d}\;; \qquad
\frac{\partial \psi_1}{\partial \xi_{ab}} = 2 \xi^{ab}; \qquad 
\frac{\partial \nu}{\partial \xi_{ab}} = \xi^a \xi^b - \frac{\mu}{d} g^{ab}\;;
\]

\[
\frac{\partial \ell^{a}}{\partial \xi_{bc}}= g^{a(b} \xi^{c)} - \frac{\xi^a g^{bc}}{d}\;;
\] 

\[
\frac{\partial^2 (\ell\cdot\ell)}{\partial \xi_{ab}}= g^{a(c} \xi^{d)} \xi^b + g^{b(c} \xi^{d)} \xi^a - \frac{2}{d}\left(\xi^c \xi^d g^{ab} + \xi^a \xi^b g^{cd}\right) + \frac{2\mu}{d} g^{ab}g^{cd}\;;
\]

\[
\xi^{a}\frac{\partial \psi_2}{\partial \xi^a} = 2 \psi_{2}\;; \qquad g^{ab}\frac{\partial^2 \psi_2}{\partial \xi^a\partial \xi^b} = 2 \psi_{1}\;;
\]

\[
\xi^{a}\frac{\partial \psi_3}{\partial \xi^a} = 4\psi_{3}\;; \qquad g^{ab}\frac{\partial^2 \psi_3}{\partial \xi^a\partial \xi^b} = 8\psi_{2}.
\]

\subsection{... second order $A^{abc}$}

\[
\frac{\partial \psi_1}{\partial \xi_{ab}}  = 2 \xi^{ab}\;; \qquad
\frac{\partial \psi_2}{\partial \xi_{ab}}  = 2 \left(\xi^{(a} \ell^{b)} - \frac{\nu}{d} g^{ab}\right)\;; \qquad
\frac{\partial \psi_3}{\partial \xi_{ab}}  = 2 \nu \left(\xi^a \xi^b - \frac{\mu}{d} g^{ab}\right)\;;
\]

\begin{eqnarray*}
\frac{\partial^2 \psi_1}{\partial \xi_a \partial \xi_{bc}} &=& 0\;; \\
\frac{\partial^2 \psi_2}{\partial \xi_a \partial \xi_{bc}} &=& 2\left(g^{a(b} \ell^{c)}+ \xi^{(b} \xi^{c)a} - \frac{2}{d} \ell^{a}  g^{bc}\right)\;; \\
\frac{\partial^2 \psi_3}{\partial \xi_a \partial \xi_{bc}} &=& 4\nu\left(g^{a(b} \xi^{c)} - \frac{1}{d}\xi^a g^{bc}\right) + 4\ell^a \left(\xi^b \xi^c - \frac{\mu}{d} g^{bc}\right).
\end{eqnarray*}
\nocite{*}

\bibliographystyle{ieeetr}
\bibliography{biblio}

\begin{thebibliography}{10}

\bibitem{geroch1990dissipative}
R.~Geroch and L.~Lindblom, ``Dissipative relativistic fluid theories of
  divergence type,'' {\em Physical Review D}, vol.~41, no.~6, p.~1855, 1990.

\bibitem{pennisi87}
S.~Pennisi, ``Some considerations on a non linear approach to extended
  thermodynamics,'' {\em Symposium of Kinetic Theory and Extended
  Thermodynamics}, Bologna, 1987.

\bibitem{VanRiper79}
K.~A. Van~Riper, ``General relativistic hydrodynamics and the adiabatic
  collapse of stellar cores,'' {\em Astrophysical Journal, Part 1, vol. 232, p.
  558-571}, 1979.

\bibitem{Zel'dovich71}
Y.~B. Zel'dovich and I.~D. Novikov, ``Relativistic astrophysics,'' {\em vol 1.
  Stars and Relativity}, Chicago: The University of Chicago Press, 1971.

\bibitem{Peebles80}
P.~J.~E. Peebles, ``The large-scale structure of the universe,'' {\em
  Princeton, N.J.: Princeton University Press.}, 1980.

\bibitem{anile1989relativistic}
A.~Anile, {\em Relativistic Fluids and Magneto-fluids: With Applications in
  Astrophysics and Plasma Physics}.
\newblock Cambridge monographs on mathematical physics, Cambridge University
  Press, 1989.

\bibitem{Barz85}
H.~W. Barz, L.~P. Csernai, B.~Kampfer, and B.~Luk\'acs, ``Stability of
  detonation fronts leading to quark-gluon plasma,'' {\em Phys. Rev. D},
  vol.~32, pp.~115--122, Jul 1985.

\bibitem{CLARE1986177}
R.~Clare and D.~Strottman, ``Relativistic hydrodynamics and heavy ion
  reactions,'' {\em Physics Reports}, vol.~141, no.~4, pp.~177 -- 280, 1986.

\bibitem{Jaiswal16}
A.~Jaiswal and V.~Roy, ``Relativistic hydrodynamics in heavy-ion collisions:
  General aspects and recent developments,'' {\em Advances in High Energy
  Physics, vol. 2016}, Article ID 9623034, 39 pages, doi:10.1155/2016/9623034,
  2016.

\bibitem{mukhanov2005physical}
V.~Mukhanov, {\em Physical Foundations of Cosmology}.
\newblock Cambridge University Press, 2005.

\bibitem{Lichne65}
A.~Lichnerowicz, ``Théorèmes d’existence et d'unicité pour un fluide
  thermodynamique relativiste,'' {\em C. R. Acad. Sci. Paris 260 (1965)
  3291-3295}, 1965.

\bibitem{Lichne66}
A.~Lichnerowicz, ``Etude mathématique des fluides thermodynamiques
  relativístes,'' {\em Comm. Math. Phys. 1 (1966) 328-373}, 1966.

\bibitem{Betz:2009zz}
B.~Betz, D.~Henkel, and D.~H. Rischke, ``{Complete second-order dissipative
  fluid dynamics},'' {\em J. Phys.}, vol.~G36, p.~064029, 2009.

\bibitem{Van:2011yn}
P.~Van and T.~S. Biro, ``{First order and stable relativistic dissipative
  hydrodynamics},'' {\em Phys. Lett.}, vol.~B709, pp.~106--110, 2012.

\bibitem{calzetta2010linking}
E.~Calzetta and J.~Peralta-Ramos, ``Linking the hydrodynamic and kinetic
  description of a dissipative relativistic conformal theory,'' {\em Physical
  Review D}, vol.~82, no.~10, p.~106003, 2010.

\bibitem{PhysRevLett.105.162501}
G.~S. Denicol, T.~Koide, and D.~H. Rischke, ``Dissipative relativistic fluid
  dynamics: A new way to derive the equations of motion from kinetic theory,''
  {\em Phys. Rev. Lett.}, vol.~105, p.~162501, Oct 2010.

\bibitem{0305-4470-28-23-033}
G.~B. Nagy and O.~A. Reula, ``On the causality of a dilute gas as a dissipative
  relativistic fluid theory of divergence type,'' {\em Journal of Physics A:
  Mathematical and General}, vol.~28, no.~23, p.~6943, 1995.

\bibitem{geroch1991causal}
R.~Geroch and L.~Lindblom, ``Causal theories of dissipative relativistic
  fluids,'' {\em Annals of Physics}, vol.~207, no.~2, pp.~394--416, 1991.

\bibitem{Hiscock:1985zz}
W.~A. Hiscock and L.~Lindblom, ``{Generic instabilities in first-order
  dissipative relativistic fluid theories},'' {\em Phys. Rev.}, vol.~D31,
  pp.~725--733, 1985.

\bibitem{nagy1997exponential}
G.~B. Nagy, O.~E. Ortiz, and O.~A. Reula, ``Exponential decay rates in
  quasi-linear hyperbolic heat conduction,'' {\em Journal of Non-Equilibrium
  Thermodynamics}, vol.~22, no.~3, pp.~248--259, 1997.

\bibitem{kreiss1997global}
H.-O. Kreiss, G.~B. Nagy, O.~E. Ortiz, and O.~A. Reula, ``Global existence and
  exponential decay for hyperbolic dissipative relativistic fluid theories,''
  {\em Journal of Mathematical Physics}, vol.~38, no.~10, pp.~5272--5279, 1997.

\bibitem{landau2013fluid}
L.~Landau and E.~Lifshitz, {\em Fluid Mechanics}.
\newblock No.~v. 6, Elsevier Science, 2013.

\bibitem{geroch2001hyperbolic}
R.~Geroch, ``On hyperbolic "theories" of relativistic dissipative fluids,''
  {\em arXiv preprint gr-qc/0103112}, 2001.

\bibitem{liu1986relativistic}
I.-S. Liu, I.~M{\"u}ller, and T.~Ruggeri, ``Relativistic thermodynamics of
  gases,'' {\em Annals of Physics}, vol.~169, no.~1, pp.~191--219, 1986.

\bibitem{muller2013extended}
I.~M{\"u}ller and T.~Ruggeri, {\em Extended thermodynamics}, vol.~37.
\newblock Springer Science \& Business Media, 2013.

\bibitem{geroch1995relativistic}
R.~Geroch, ``Relativistic theories of dissipative fluids,'' {\em Journal of
  Mathematical Physics}, vol.~36, no.~8, pp.~4226--4241, 1995.

\bibitem{bhattacharyya2008nonlinear}
S.~Bhattacharyya, S.~Minwalla, V.~E. Hubeny, and M.~Rangamani, ``Nonlinear
  fluid dynamics from gravity,'' {\em Journal of High Energy Physics},
  vol.~2008, no.~02, p.~045, 2008.

\bibitem{boffetta12}
G.~Boffetta and R.~E. Ecke, ``Two-dimensional turbulence,'' {\em Annual Review
  of Fluid Mechanics}, vol.~44, no.~1, pp.~427--451, 2012.

\bibitem{Bhattacharyya:2008mz}
S.~Bhattacharyya, R.~Loganayagam, I.~Mandal, S.~Minwalla, and A.~Sharma,
  ``{Conformal Nonlinear Fluid Dynamics from Gravity in Arbitrary
  Dimensions},'' {\em JHEP}, vol.~12, p.~116, 2008.

\bibitem{Rangamani:2009xk}
M.~Rangamani, ``{Gravity and Hydrodynamics: Lectures on the fluid-gravity
  correspondence},'' {\em Class. Quant. Grav.}, vol.~26, p.~224003, 2009.

\bibitem{Ambrosetti:2008mt}
N.~Ambrosetti, J.~Charbonneau, and S.~Weinfurtner, ``{The Fluid/gravity
  correspondence: Lectures notes from the 2008 Summer School on Particles,
  Fields, and Strings},'' in {\em {6th Summer School on Particles, Fields and
  Strings Vancouver, British Columbia, Canada, July 22-August 1, 2008}}, 2008.

\bibitem{Bredberg2012}
I.~Bredberg, C.~Keeler, V.~Lysov, and A.~Strominger, ``From navier-stokes to
  einstein,'' {\em Journal of High Energy Physics}, vol.~2012, p.~146, Jul
  2012.

\bibitem{1126-6708-2008-05-087}
R.~Loganayagam, ``Entropy current in conformal hydrodynamics,'' {\em Journal of
  High Energy Physics}, vol.~2008, no.~05, p.~087, 2008.

\bibitem{geroch1996partial}
R.~Geroch, ``Partial differential equations of physics,'' {\em General
  Relativity, Aberdeen, Scotland}, pp.~19--60, 1996.

\bibitem{0264-9381-15-3-015}
E.~Calzetta, ``Relativistic fluctuating hydrodynamics,'' {\em Classical and
  Quantum Gravity}, vol.~15, no.~3, p.~653, 1998.

\bibitem{rezzolla2013relativistic}
L.~Rezzolla and O.~Zanotti, {\em Relativistic Hydrodynamics}.
\newblock OUP Oxford, 2013.

\bibitem{doi:10.1119/1.1336839}
R.~Baierlein, ``The elusive chemical potential,'' {\em American Journal of
  Physics}, vol.~69, no.~4, pp.~423--434, 2001.

\bibitem{wald2010general}
R.~Wald, {\em General Relativity}.
\newblock University of Chicago Press, 2010.

\bibitem{Tolman:1930ona}
R.~Tolman and P.~Ehrenfest, ``{Temperature Equilibrium in a Static
  Gravitational Field},'' {\em Phys. Rev.}, vol.~36, no.~12, pp.~1791--1798,
  1930.

\bibitem{Smerlak2012}
M.~Smerlak, ``On the inertia of heat,'' {\em The European Physical Journal
  Plus}, vol.~127, p.~72, Jul 2012.

\bibitem{kreiss2004initial}
H.~Kreiss and J.~Lorenz, {\em Initial-Boundary Value Problems and the
  Navier-Stokes Equation:}.
\newblock Classics in Applied Mathematics, Society for Industrial and Applied
  Mathematics, 2004.

\bibitem{BALAZS1965222}
N.~Balazs and J.~Dawson, ``On thermodynamic equilibrium in a gravitational
  field,'' {\em Physica}, vol.~31, no.~2, pp.~222 -- 232, 1965.

\bibitem{hiscock1983stability}
W.~A. Hiscock and L.~Lindblom, ``Stability and causality in dissipative
  relativistic fluids,'' {\em Annals of Physics}, vol.~151, no.~2,
  pp.~466--496, 1983.

\bibitem{baier2008relativistic}
R.~Baier, P.~Romatschke, D.~T. Son, A.~O. Starinets, and M.~A. Stephanov,
  ``Relativistic viscous hydrodynamics, conformal invariance, and holography,''
  {\em Journal of High Energy Physics}, vol.~2008, no.~04, p.~100, 2008.

\bibitem{0264-9381-27-2-025006}
P.~Romatschke, ``Relativistic viscous fluid dynamics and non-equilibrium
  entropy,'' {\em Classical and Quantum Gravity}, vol.~27, no.~2, p.~025006,
  2010.

\bibitem{PhysRevD.81.114039}
S.~Pu, T.~Koide, and D.~H. Rischke, ``Does stability of relativistic
  dissipative fluid dynamics imply causality?,'' {\em Phys. Rev. D}, vol.~81,
  p.~114039, Jun 2010.

\bibitem{0954-3899-35-11-115102}
G.~S. Denicol, T.~Kodama, T.~Koide, and P.~Mota, ``Stability and causality in
  relativistic dissipative hydrodynamics,'' {\em Journal of Physics G: Nuclear
  and Particle Physics}, vol.~35, no.~11, p.~115102, 2008.

\bibitem{friedrichs1954symmetric}
K.~O. Friedrichs, ``Symmetric hyperbolic linear differential equations,'' {\em
  Communications on pure and applied Mathematics}, vol.~7, no.~2, pp.~345--392,
  1954.

\bibitem{friedrichs1971systems}
K.~O. Friedrichs and P.~D. Lax, ``Systems of conservation equations with a
  convex extension,'' {\em Proceedings of the National Academy of Sciences},
  vol.~68, no.~8, pp.~1686--1688, 1971.

\bibitem{Bemfica:2017wps}
F.~S. Bemfica, M.~M. Disconzi, and J.~Noronha, ``{Causality and existence of
  solutions of relativistic viscous fluid dynamics with gravity},'' 2017.

\end{thebibliography}

\end{document}